\documentclass[twocolumn,superscriptaddress]{revtex4-2}
\usepackage[english]{babel}
\usepackage{booktabs} 
\usepackage[colorinlistoftodos, color=green!40, prependcaption]{todonotes}
\usepackage{amsthm}
\DeclareUnicodeCharacter{2212}{-}
\usepackage{mathtools}
\usepackage{physics}
\usepackage{xcolor}
\usepackage{graphicx}
\usepackage{comment}
\usepackage[normalem]{ulem}
\usepackage[left=23mm,right=13mm,top=35mm,columnsep=15pt]{geometry} 
\usepackage{lipsum}
\usepackage{placeins}
\usepackage{csquotes}
\usepackage[pdftex, pdftitle={Article}, pdfauthor={Author}]{hyperref}
\usepackage[dvipsnames]{xcolor}
\usepackage{bm}
\usepackage{siunitx}
\usepackage{xr}
\usepackage{comment}
\usepackage{soul}

\addto\captionsenglish{}

\begin{document}

\title{ Transition Waves in Mechanical Metamaterials with Neighbor-Programmable Energy Landscapes}
\author{E.~Duval}
\thanks{E.~Duval and G.~Risso contributed equally to this work.}
\affiliation{Laboratoire d'Acoustique de l'Universit\'e du Mans (LAUM), UMR 6613, Institut d'Acoustique - Graduate School (IA-GS), CNRS, Le Mans Universit\'e, France}
\author{G.~Risso}
\thanks{E.~Duval and G.~Risso contributed equally to this work.}
\affiliation{School of Engineering and Applied Sciences, Harvard University, Cambridge,	Massachusetts 02138, USA}
\author{A.~Zhang}
\affiliation{School of Engineering and Applied Sciences, Harvard University, Cambridge,	Massachusetts 02138, USA}
\affiliation{Department of Mechanical \& Industrial Engineering, University of Toronto, Toronto, ON M5S 3G8, Canada}
\author{V.~Tournat}
\affiliation{Laboratoire d'Acoustique de l'Universit\'e du Mans (LAUM), UMR 6613, Institut d'Acoustique - Graduate School (IA-GS), CNRS, Le Mans Universit\'e, France}
\affiliation{School of Engineering and Applied Sciences, Harvard University, Cambridge,	Massachusetts 02138, USA}
\author{K.~Bertoldi}
\affiliation{School of Engineering and Applied Sciences, Harvard University, Cambridge,	Massachusetts 02138, USA}

\begin{abstract}
Transition waves in mechanical metamaterials manifest themselves as propagating interfaces between different stable states in lattices composed of arrays of coupled, intrinsically bistable elements. Here, we show experimentally and numerically that arrays of elastic unit cells that are individually monostable, yet whose energy landscapes can be programmed through interactions with neighboring units, provide a rich and largely unexplored platform for transition wave propagation. We implement this concept by designing a unit cell comprising a von Mises truss supported by two vertical elastic beams. In one-dimensional arrays of such units, we demonstrate that each cell's energy landscape can change from monostable to bistable depending on the state of its neighbors.
This neighbor-programmable energy landscape enables the controlled initiation and propagation of transition waves, giving rise to highly discrete, directionally unbiased, domino-like wave propagation. Experiments and numerical simulations show that the existence and speed of the waves are governed by geometric design and mass distribution. Our results establish neighboring effects as a distinct mechanism for transition wave propagation, expanding the design space of mechanical metamaterials beyond architectures that rely on intrinsically multistable building blocks.

\end{abstract}

\keywords{}
\maketitle

In mechanical metamaterials composed of arrays of bistable elements, the switching of a unit from one equilibrium configuration to another can trigger neighboring units to undergo the same transition. This sequential process gives rise to a transition wave, a moving front that propagates through the lattice and separates regions in distinct states. As a result, transition wave propagation in mechanical metamaterials has traditionally relied on two essential ingredients: bistability at the unit-cell level and sufficiently strong coupling between neighboring elements. 

A wide variety of bistable architectures has been explored to support such behavior, including beams~\cite{Nadkarni_2014,Nadkarni_2016,raney2016stable,Hwang2018,jin2020guided,Bonthron2025,Bronstein2025}, rotating squares with embedded magnets~\cite{Yasuda_2020,yasuda2023} and shells~\cite{Vasios2021}. Equally critical is the mechanism that couples adjacent units. Different interaction strategies have been implemented, including magnetic forces~\cite{Nadkarni_2014,Bilal_2017,Ray2024,Pal2023,Hwang2018}, elastic springs~\cite{raney2016stable,Frazier2017}, or continuum beams \cite{jin2026transition}, all aimed at transmitting the switching event from one unit to the next and thereby sustaining wave propagation.

Here, we introduce a new platform for transition wave propagation: mechanical metamaterials whose energy landscapes can be programmed through interactions with neighboring units~\cite{Wu2023,RISSO2025102393,WU2026,Wu2026_advmat, Hasan2026}. In these systems, the effective energy landscape of each unit is not purely intrinsic but instead depends on the deformation state of adjacent elements.
We show that such neighbor-mediated interactions naturally lead to the emergence of transition waves. Focusing on a one-dimensional array of von Mises truss units connected by flexible beams, we combine experiments and numerical simulations to demonstrate the propagation of transition waves and to examine how the wave speed depends on the geometric parameter defining the unit cell. Together, these findings establish neighbor-induced interactions as a new mechanism for transition wave propagation in mechanical metamaterials.

\begin{figure*}[!hpt]
\begin{center}
\includegraphics[width = 1.99\columnwidth]{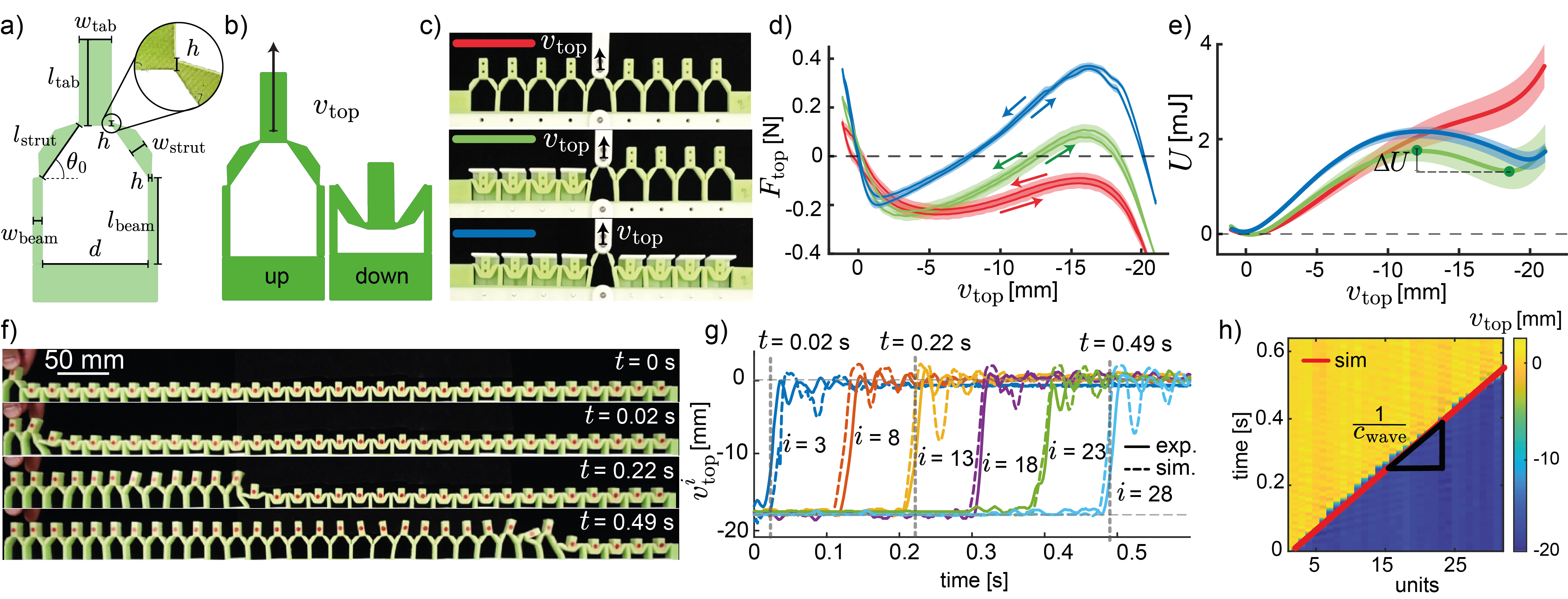} 
\caption{\label{fig:fig1}\textbf{Transition waves in mechanical metamaterials with neighbor-programmable energy landscapes.} 
\textbf{a)} Schematic of the unit cell. 
\textbf{b)} The von Mises truss exhibits two distinct stable equilibrium states: the natural, stress-free \emph{up state} and the inverted \emph{down state}.  
\textbf{c)} Experimental snapshots of the mechanical characterization tests. A displacement is applied to the pull tab of the central unit while (i) all neighboring units are in the up state (red), (ii) the left neighbors are in the down state and the right neighbors in the up state (green), and (iii) all neighbors are in the down state (blue). 
\textbf{d)} Experimentally measured force–displacement curves. 
\textbf{e)} Corresponding energy landscapes. 
\textbf{f)} Sequential snapshots showing a transition wave propagating along the 32-unit metamaterial at four representative time points. 
\textbf{g)} Experimental (solid lines) and numerical (dashed lines) results for the temporal evolution of the vertical displacement $v^i_{\text{top}}$ of selected units $i$. 
\textbf{h)} Spatiotemporal displacement diagram. Supplementary Video~1 provides the corresponding experimental and numerical footage.}
\vspace{-15pt}
    \end{center}
\end{figure*}

Our structure is composed of a one-dimensional array of repeated unit cells of out-of-plane thickness $t_{\text{out}}$. Each unit cell contains a von Mises truss joined at its base to two vertical elastic beams of length $l_{\text{beam}}$ and width $w_{\text{beam}}$, spaced by a distance $d$ (Fig.~\ref{fig:fig1}a). The von Mises truss consists of two identical struts of length $l_{\text{strut}}$ and width $w_{\text{strut}}$, arranged symmetrically at an angle $\theta_0$ relative to the horizontal. These struts connect to both the vertical beams and a rectangular pull tab of width $w_{\text{tab}}$ and height $l_{\text{tab}}$---included to facilitate loading---through tapered hinges of width $h$. As shown in Fig.~\ref{fig:fig1}b, the von Mises truss may exhibit two distinct stable equilibrium states: the natural stress-free \emph{up state} and the inverted \emph{down state}. Throughout this study, we fix the geometric parameters to $t_{\text{out}} = 10$~mm, $w_{\text{strut}} = 3$~mm, $l_{\text{strut}} = 11$~mm, $w_{\text{tab}} = 6$~mm, $d = 18.4$~mm, and $l_{\text{beam}} = 15$~mm, resulting in $\theta_0=\arccos[(d - w_{\text{tab}})/(2l_{\text{strut}})]= 55.7^{\circ}$ and investigate the influence of $h$ and $w_{\text{beam}}$.

The structures are fabricated via molding from a low-viscosity silicone elastomer (Zhermack Elite Double 32A, Young’s modulus $E = 1.3$~MPa, Poisson’s ratio $\nu = 0.5$, density $\rho = 1200$~kg/m$^{3}$), which allows for large and reversible deformations (see Supplementary Material, Section~B for details). To characterize the mechanical response, we perform quasi-static tests on 9-unit samples. In these tests, the base of the structure is rigidly attached to a stiff substrate, while the pull tab of the central unit is connected to a translational stage (Fig.~\ref{fig:fig1}c). A vertical displacement $v_{\text{top}}$ is imposed on the central unit at a constant speed of $2$~mm/s, starting from the stress-free \emph{up state} (see Supplementary Material, Section~C for details). In Figs.~\ref{fig:fig1}c and~\ref{fig:fig1}d, we focus on a sample with $h = 0.7$~mm and $w_{\text{beam}} = 1.4$, and report the evolution of the measured reaction force $F_{\text{top}}$ and corresponding potential energy $U$ as a function of $v_{\text{top}}$, respectively. The potential energy is defined such that $\partial U / \partial v_{\text{top}} = -F_{\text{top}}$, and is obtained by numerically integrating the measured force--displacement data, with the reference chosen so that $U(v_{\text{top}} = 0) = 0$.
 When all unit cells are in their stress-free \emph{up state}, pushing the von Mises truss of the central unit downward eventually causes it to snap into the inverted state, as indicated by the recorded non-monotonic force–displacement response (red curve in Fig.~\ref{fig:fig1}d). $F_{\text{top}}$ remains always negative and $U$ displays a single minimum at $v_{top}=0$, indicating that only the \emph{up state} is stable.  In contrast, when all non-actuated von Mises trusses are kept in their \emph{down state} using tabs, $F_{\text{top}}$ crosses zero and $U$ display two minima at $v_{top}$=0 and $-18.2$mm separated by an energy barrier, showing that the central von Mises truss becomes bistable (blue curve in Figs.~\ref{fig:fig1}d-e). This occurs because a von Mises truss in the \emph{down state} forces the connected vertical beams to bend outward, restricting the horizontal motion of neighboring trusses and thereby promoting their bistability. 
 Finally, when only the von Mises trusses on one side of the central unit are kept in their \emph{down state} using tabs, the central von Mises truss becomes marginally bistable, as the associated energy barrier is greatly reduced (green curve in Figs.~\ref{fig:fig1}d-e). These observations demonstrate that, for the considered structure, the state of neighboring units strongly influences the energy landscape of the central one.

Since each unit cell shifts from monostable to marginally bistable to fully bistable depending on the states of its neighbors, the structure provides a natural platform for the propagation of transition waves. In Fig.~\ref{fig:fig1}f, we consider a $32$-unit metamaterial with $h = 0.7$~mm, and $w_{\text{beam}}=1.4$ mm. We initialize the system with all von Mises trusses in their \emph{down state}, and the interactions with neighboring units ensure that this configuration is stable. However, when the von Mises truss of one unit is lifted, the von Mises truss of the neighboring unit becomes marginally bistable, and the perturbation introduced by this change in state is sufficient to make its down state unstable and trigger a transition to the up configuration. This, in turn, renders the next unit marginally bistable, triggering a sequential cascading response (see Fig.~\ref{fig:fig1}f and Supplementary Video 1). As a result, a domino-like transition wave propagates through the structure, with units snapping up to their up state one after another. To quantify this behavior, we marked the pull tab of each unit with a red dot, recorded the experiments with a high-speed camera (Sony RX100V recording at 480 fps), and extracted the vertical displacement of the markers with a custom MATLAB script. In Fig.~\ref{fig:fig1}g, we report the evolution of the tracked displacement $v_{\text{top}}$ of selected unit cells as a function of time. We find that the wave propagates in a discrete, stepwise manner, with a width of a single unit. Moreover, the corresponding spatiotemporal plot in Fig.~\ref{fig:fig1}h shows that the wave propagates at a constant speed measured to be $c_\text{wave}=54.1$~units/s. 

\begin{figure}
\begin{center}
\includegraphics[width = 0.99\columnwidth]{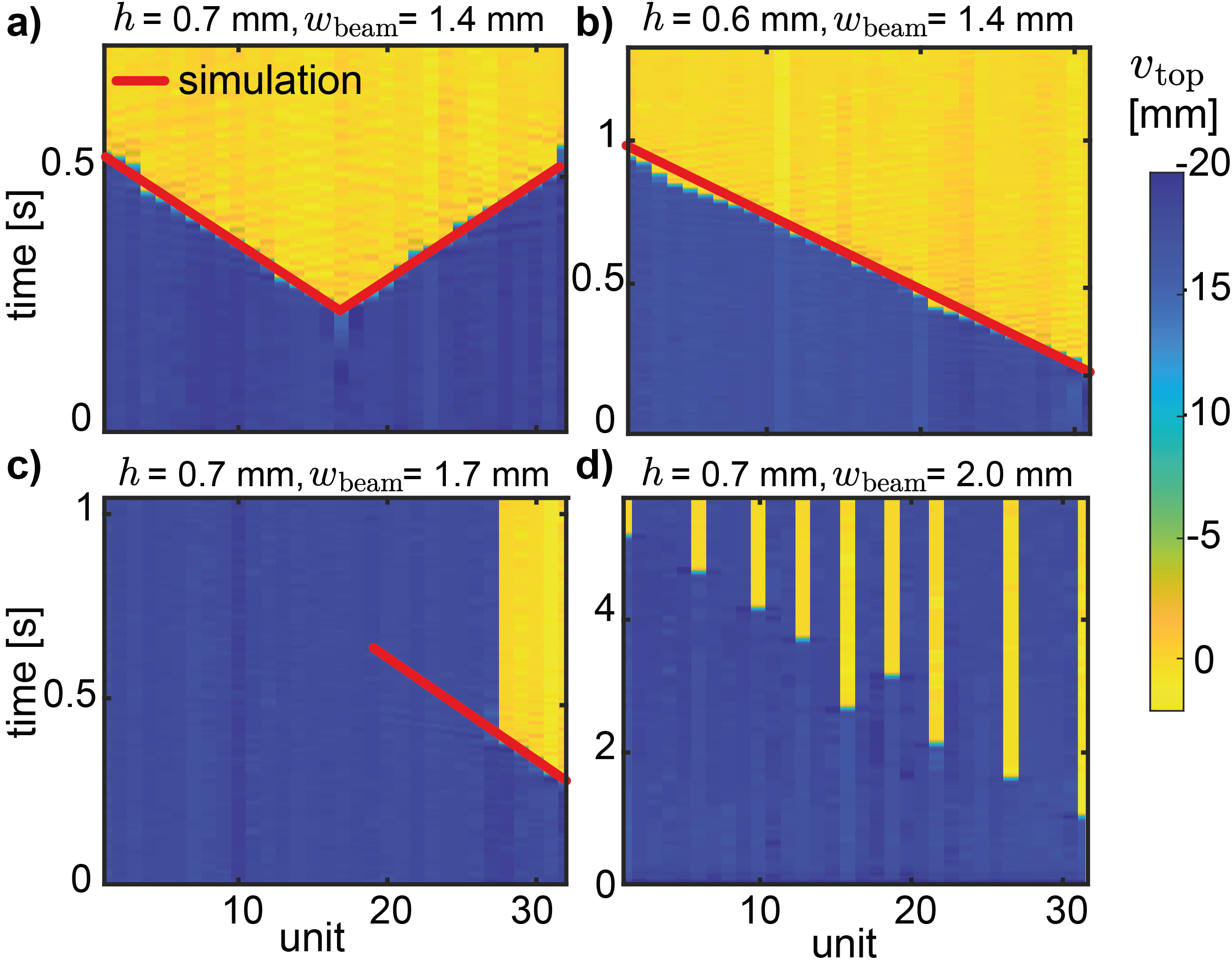} 
\caption{\label{fig:fig3} \textbf{Effect of geometry on wave propagation.} Spatiotemporal displacement diagrams for the four geometries investigated, overlaid with the corresponding numerical predictions (red lines). Supplementary Video~2 presents the associated experimental recordings.}
\vspace{-15pt}
\end{center}
\end{figure}

The transition waves propagate robustly through the sample in both directions, meaning the wave can travel along the chain in either spatial direction (see Supplementary Video 2 for experimental results). 
When the central unit is pulled upward, two transition waves are generated that travel outward from the actuation site at the same speed, $c_{\text{wave}} = 54.1$~units/s (Fig.~\ref{fig:fig3}a). The geometry of the unit cells strongly affects the wave-propagation behavior of the metamaterial. For instance, when $h$ is reduced to 0.6~mm while the beam width $w_{\text{beam}}$ remains at 1.4~mm, similar to the structure shown in Fig.~\ref{fig:fig1}, where each unit cell shifts from monostable, to marginally bistable and to fully bistable depending on the states of its neighbors (Fig.~S6-S10). Under these conditions, transition waves still propagate, but at the reduced speed of $c_{\text{wave}} = 40.1$~units/s.
In contrast, when $h = 0.7$~mm is held fixed and $w_{\text{beam}}$ is progressively increased, the wave initially propagates across only a few units at the speed of $c_{\text{wave}} = 37.2 $~units/s (Fig.~\ref{fig:fig3}c and Fig.~S10) and ultimately fails to initiate at all (Fig.~\ref{fig:fig3}d). This occurs because larger values of $w_{\text{beam}}$ increase the energy barrier associated with switching from the down to the up states, making a unit cell bistable even when only the von Mises trusses on one side of the central unit are in their down state (Fig.~S7-S10). As this barrier grows, the unit cell eventually becomes bistable for any configuration of its neighboring units (Fig.~S8-S10), preventing wave propagation entirely.

\begin{figure}[!hpt]
\begin{center}
\includegraphics[width = 0.99\columnwidth]{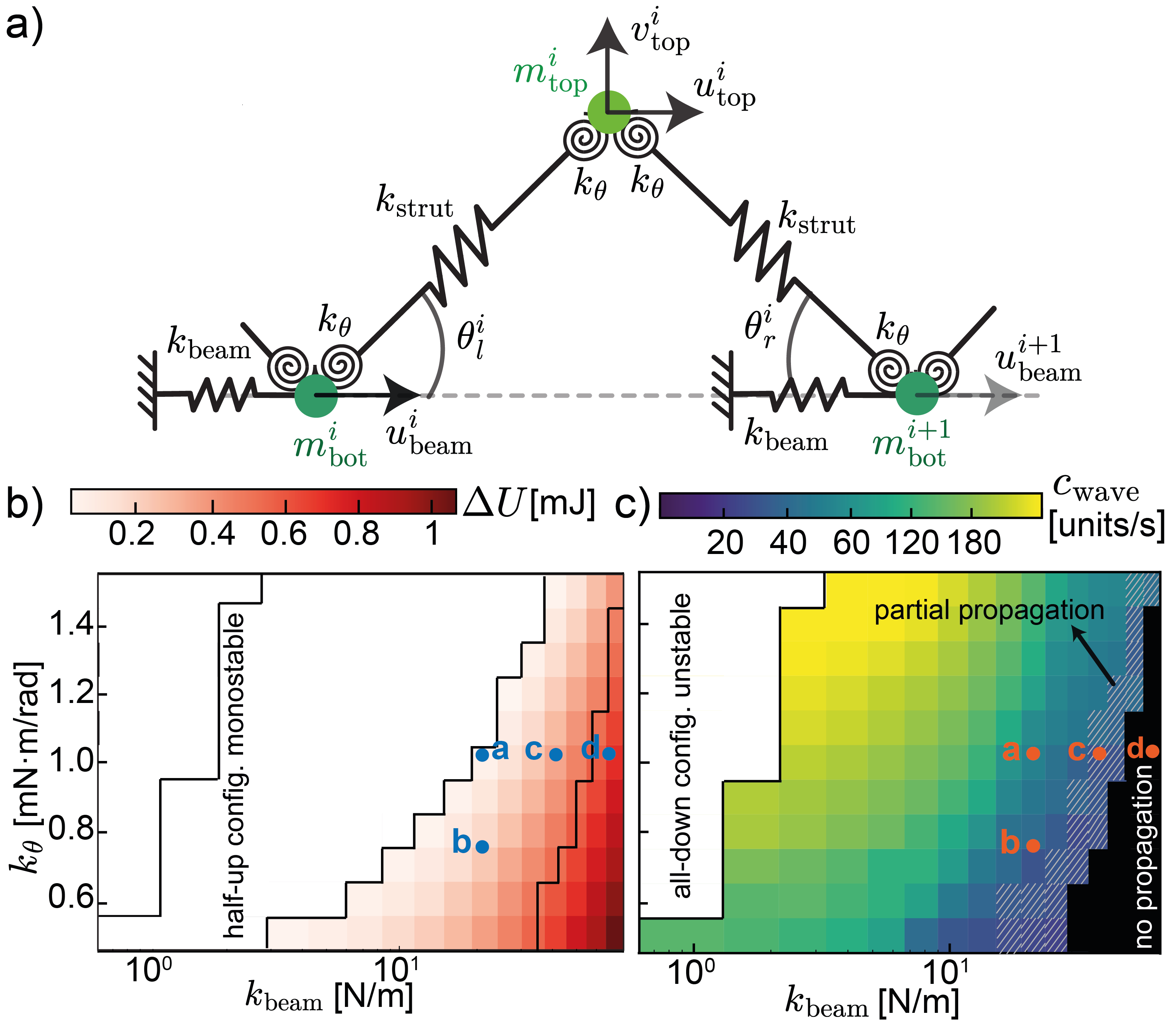} 
\caption{\label{fig:fig3a} \textbf{Systematic exploration of the design space.} 
\textbf{a)} Schematic of the mass–spring model. 
\textbf{b)} Numerically predicted evolution of the energy barrier $\Delta U$ as a function of $k_\theta$ and $k_\text{beam}$. 
\textbf{c)} Numerically predicted evolution of the wave speed $c_{\text{wave}}$ as a function of $k_\theta$ and $k_\text{beam}$. Red and blue markers indicate the four experimental structures considered in Fig.~\ref{fig:fig3}, with labels (a–d) corresponding to the panels shown there.}
\vspace{-15pt}
\end{center}
\end{figure}

To systematically investigate how geometry influences the system’s dynamic response, we introduce a spring--mass model that captures the primary deformation mechanisms observed in the experiments. As illustrated in Fig.~\ref{fig:fig3a}a, each unit cell is represented by two concentrated masses connected through a network of linear axial and torsional springs.  A concentrated mass $m^i_{\text{bot}}$ with a single degree of freedom (horizontal displacement $u^i_{\text{beam}}$) is located at the top of the left vertical beam of the $i$-th unit cell and represents the mass of the vertical beam together with halves of the masses of the two adjacent inclined struts. The second concentrated mass $m^i_{\text{top}}$ is located at the bottom of the $i^\text{th}$ vertical pull tab and has two degrees of freedom (horizontal displacement $u^i_{\text{top}}$ and vertical displacement $v^i_{\text{top}}$). It represents the mass of the vertical tab located at the top of each unit cell and half of the mass of the adjacent inclined struts. As for the springs, the $i$-th unit cell comprises: (i) two axial springs of stiffness $k_\text{strut}$, representing the extension and compression of the inclined struts under loading; (ii) one axial spring of stiffness $k_\text{beam}$, representing the transverse deformation of the vertical beams under lateral loading; and (iii) four torsional springs of stiffness $k_\theta$, modeling the response of the thin ligaments located at the junctions between the struts and the pull bars or the vertical base beams. $k_\text{strut}$ and $k_\text{beam}$ are determined as $k_\text{beam}= 3 E t_\text{out} (2w_\text{beam})^3/(12l_\text{beam}^3)$  and $k_\text{strut} = E t_\text{out} w_\text{strut}/l_\text{strut}$, while $k_{\theta}$ is fitted to quasi-static experimental data (see Supplementary Materials Section~D3 for details). The equations of motion for this spring--mass system can be written as:

\begin{subequations}

\label{eq:newton_laws_main}
\begin{equation}
\label{eq:newton_laws_main_top}
m_\text{top}^i\mathbf{a}_{\text{top}}
= - \mathbf{F}^i_{\text{strut},l} - \mathbf{F}^i_{\text{strut},r}
  + \mathbf{F}^{i}_{\theta,\text{top}} + c_\text{top} \dot{u}_\text{top}^i
\end{equation}
\begin{equation}
\label{eq:newton_laws_main_beam}
\begin{aligned}
m^i_\text{bot}\,\ddot{u}_\text{beam}^i
&= \left(\mathbf{F}^i_{\text{strut},l} + \mathbf{F}^i_{\text{strut},r}
        + \mathbf{F}^i_{\text{beam}}\right)\cdot\mathbf{e}_x  \\
&\quad + F^i_{\theta,\text{bottom}} + c_\text{beam} \dot{u}_\text{beam}^i
\end{aligned}
\end{equation}
\end{subequations}
where $\mathbf{a}_{\text{top}}=[\ddot{u}_\text{top}^i \,\ddot{v}_\text{top}^i]$ and $\mathbf{F}^i_{\text{strut},l}$ and $\mathbf{F}^i_{\text{strut},r}$ represent the axial restoring forces generated by the deformation of the inclined left and right struts of the von Mises truss, respectively. Further, $\mathbf{F}^i_{\text{beam}}$ denotes the horizontal force generated by $k_{\text{beam}}$ acting on $m^i_{\text{bot}}$ and $F^i_{\theta, \text{top}}$ and $\mathbf{F}^{i}_{\theta, \text{bottom}}$ represent the torque-induced forces arising from the rotational springs at the hinges acting on top and bottom masses, respectively. Finally, although elastomers exhibit complex dissipative behavior, we approximate this behavior using a linear viscous damping model with linear viscous damping coefficients $c_{\text{beam}} = 0.022~\mathrm{N\!\cdot\!s/m}$ and 
$c_{\text{top}} = 0.016~\mathrm{N\!\cdot\!s/m}$  determined by fitting our model to experimental data 
(see Supplementary Materials, Section~D1, for details). 

In our simulations, we consider a chain of $N$ unit cells and assume that at time $t = 0$ all units are in their down configurations. To trigger wave propagation, we perturb the $i$-th unit at $t = 0$ by imposing $v_{\text{top}}^{\,i}(0) = - 1~\text{mm}$, and we numerically integrate Eqs.~(\ref{eq:newton_laws_main_beam}) using the \texttt{odeint} solver in \textsc{JAX} \cite{bradbury_2018}, which employs an adaptive step-size Runge--Kutta method. As shown in Figs.~\ref{fig:fig1} and ~\ref{fig:fig3}, and Supplementary Video 1, the model accurately reproduces both the vertical displacements of individual units during propagation and the measured propagation velocities across different geometries, confirming the validity of our approach.

Next, we use the model to investigate how geometry influences the energy landscape and the propagation of transition waves and simulate the response of systems with \mbox{$k_\theta \in [0.5, 1.5]~\mathrm{mN}\cdot\mathrm{m/rad}$} and \mbox{$k_{\text{beam}} \in [1, 71]~\mathrm{N/m}$}. In Fig.~\ref{fig:fig3a}b, we report the numerically predicted evolution of the energy barrier, $\Delta U$, associated with the transition from the down to the up state for configurations in which the von Mises trusses on one side of the central unit are maintained in the down state and those on the other side in the up state (see Fig.~\ref{fig:fig1}e). 

In Fig.~\ref{fig:fig3a}c, we present the corresponding numerically predicted wave speed, $c_{\text{wave}}$. We find that for small values of $k_{\text{beam}}$, there is no energy barrier, i.e. $\Delta U \leq 0$, independently of $k_\theta$. For sufficiently small $k_{\text{beam}}$, a unit is monostable even when all other units are in their down state; in this regime, no transition wave propagation is possible (white region in Fig.~\ref{fig:fig3a}c). As $k_{\text{beam}}$ is slightly increased, a unit becomes stable when all units are in the down state, yet remains monostable when units on one side are down and those on the other side are up, so that $\Delta U$ remains $\leq 0$. In this regime, waves can propagate at very high speed, since once a neighboring unit snaps to the up state, the adjacent down unit immediately loses stability.

Upon further increasing $k_{\text{beam}}$, a unit becomes bistable both when all units are down and when units on one side are down and those on the other side are up. The magnitude of the energy barrier, $\Delta U$, increases monotonically with $k_{\text{beam}}$ and is only weakly influenced by $k_\theta$. As $\Delta U$ increases, the wave speed $c_{\text{wave}}$ decreases, and eventually wave propagation is sustained only over a limited number of units (dashed region in Fig.~\ref{fig:fig3a}c). The structure considered in Fig.~\ref{fig:fig3}c falls into this regime. For sufficiently large $k_{\text{beam}}$, a unit becomes bistable regardless of the configuration of its neighboring units, thereby suppressing transition wave propagation (black region in Fig.~\ref{fig:fig3a}c).

The results in Fig.~\ref{fig:fig3a}b indicate that $c_\text{wave}$ can be tuned by adjusting the geometry of the unit cell. While this requires fabricating new structures, the wave speed can also be tuned on demand by modifying the mass of the von Mises truss.  This strategy additionally enables spatially varying control: by selectively modifying the mass of individual units, one can locally tailor the propagation velocity. 
To demonstrate this, we consider the 32-unit structure shown in Fig.~\ref{fig:fig1} ($h = 0.7$~mm and $w_{\text{beam}} = 1.4$~mm) and investigate transition-wave propagation both experimentally and numerically as the mass of the vertical pull tab is varied. 

\begin{figure}[!hpt]
\begin{center}
\includegraphics[width = 0.99\columnwidth]{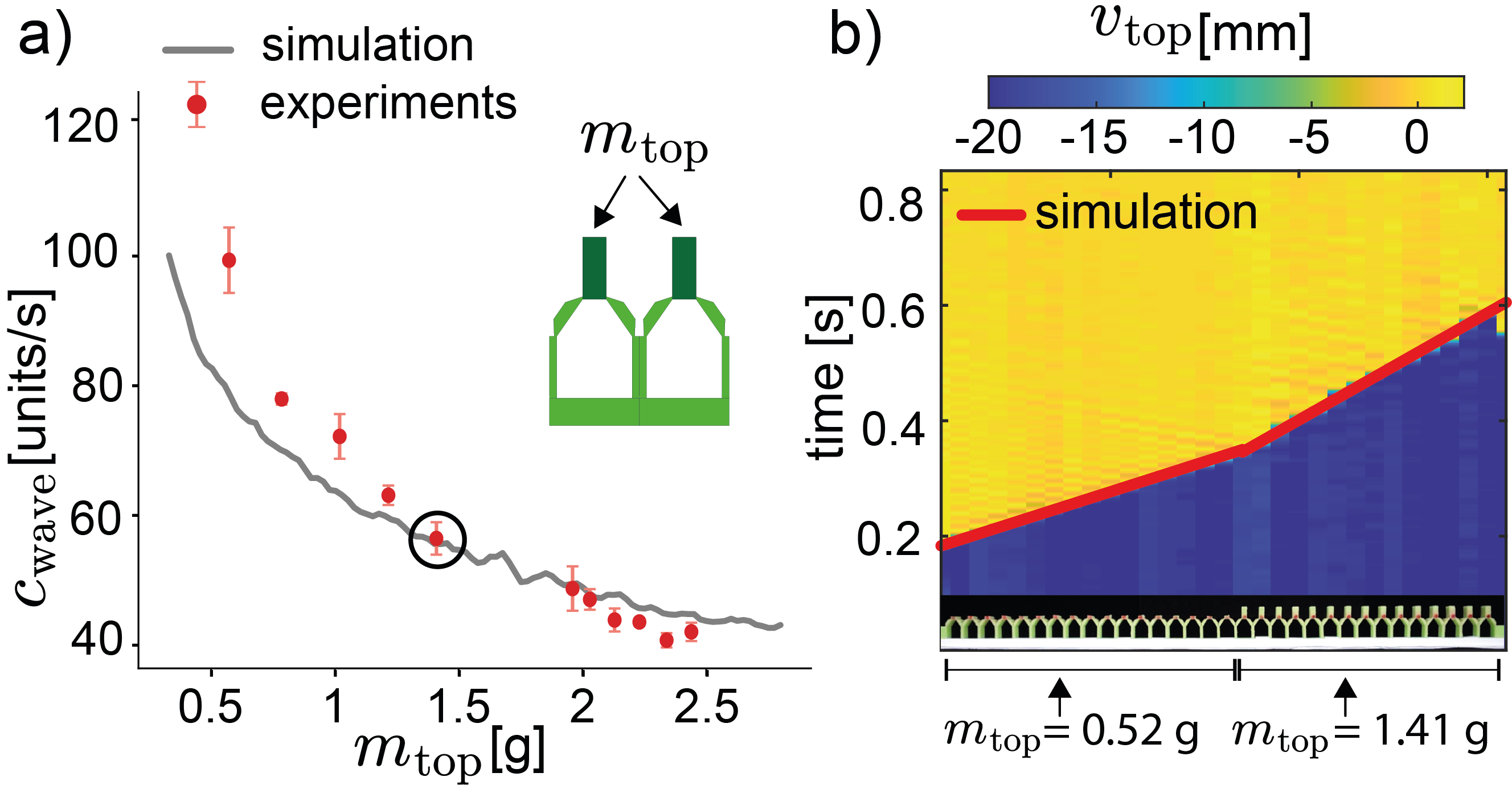} 
\caption{\label{fig:fig4} \textbf{On-demand tuning of the wave speed.} All results correspond to structures with $h = 0.7$~mm and $w = 1.4$~mm. 
\textbf{a)} Numerically predicted and experimentally measured evolution of the wave speed $c_{\text{wave}}$ as a function of the vertical pull-tab mass $m_{\text{top}}$. The reference experimental data point is highlighted with a circle. 
\textbf{b)} Spatiotemporal displacement diagram for a structure with two distinct masses: $\forall j \in [1,17],\, m^j_{\text{top}} = 0.52$~g, and $\forall j \in [18,32],\, m^j_{\text{top}} = 1.41$~g (see Supplementary Video~3).}
\vspace{-15pt}
    \end{center}
\end{figure}

Experimentally, the mass is modified either by attaching small 3D-printed weights to the pull tab or by trimming the tab to reduce its mass (see Fig.~S16), allowing access to $m_{\text{top}} \in [0.57,\, 2.44]$~g. In simulations, the mass is varied over a slightly broader range, $m_{\text{top}} \in [0.33,\, 2.8]$~g. As shown in Fig.~\ref{fig:fig4}, both experiments and simulations reveal a monotonic increase in $c_{\text{wave}}$ with decreasing $m_{\text{top}}$. In the experiments, the wave speed ranges from $92.9$~units/s for $m_{\text{top}} = 0.57$~g to $40.8$~units/s for $m_{\text{top}} = 2.44$~g.
Furthermore, decreasing $m_{\text{top}}$ on only one half of the 32-unit chain leads to a pronounced change in wave speed along the structure, from $c_{\text{wave}} = 74.9$~units/s in the left half of the structure with $m_{\text{top}} = 0.52$~g to $c_{\text{wave}} = 53.7$~units/s in the right half of the structure with $m_{\text{top}} = 1.41$~g (Fig.~\ref{fig:fig4}b and Supplementary Video 3). This result demonstrates that local tuning of the mass distribution provides a practical means of spatially controlling wave propagation. 

In summary, we have introduced a mechanical metamaterial in which transition waves arise not from intrinsically bistable unit cells, but from elements whose energy landscapes are programmed through interactions with their neighbors. Using arrays of von Mises truss units connected by flexible beams, we demonstrated experimentally and numerically that each cell can transition from monostable to bistable depending on the state of adjacent units, enabling controlled, domino-like wave propagation with tunable speed. A spring–mass model accurately captures the dynamics and clarifies how geometry and mass distribution govern both the existence and velocity of the waves, establishing a new mechanism for realizing transition waves in mechanical metamaterials. We further explored strategies to tune the wave speed on demand. While Fig.~\ref{fig:fig4} shows that this can be achieved by modifying the mass of the von Mises truss, alternative approaches are also possible. For instance, introducing base curvature alters the effective coupling between neighboring units, providing an additional means to control propagation speed (see Supplementary Materials, Section~B).

More broadly, our findings establish neighbor-induced, reprogrammable stability as a general mechanism for global reconfiguration in mechanical metamaterials. Because this concept relies solely on elastic coupling and geometric compatibility, it can be readily extended to higher-dimensional architectures \cite{Jiao2024}, opening pathways toward programmable, wave-based functionality in planar and three-dimensional metamaterial systems.
Importantly, unlike previously investigated platforms ~\cite{Nadkarni_2014,Nadkarni_2016,raney2016stable,Hwang2018,jin2020guided}, the displacement associated to state switching in our system is orthogonal to the direction of wave propagation. This distinctive feature endows the system with additional functionalities. For example, the chain could provide pulse-mitigation capabilities under impacts from above~\cite{Lakes2001, Shan2015, Morris2019}, with the number of engaged units determined by the impact geometry, while the energy absorption and post-impact configuration are governed by the units’ potential-energy landscapes and the impact characteristics. These coupled dependencies may be exploited to characterize impacts, identify objects, and localize impact events.

\hspace{0.5cm}
\section*{Data availability statement}\noindent
The data supporting the experimental results in this study are available from the corresponding author upon reasonable request. The code supporting the numerical results, including dynamical simulations, quasi-static force-displacement simulations, and postprocessing of the experimental video for tracking the metamaterial wave propagation, is openly available \cite{code}.
\hspace{0.5cm}
\section*{Acknowledgments}\noindent
G.R. acknowledges support from the Swiss National Science Foundation under Grant No. P500PT-217901. E.D. thanks the Institut d'Acoustique - Graduate School at Le Mans University for financial support (IA-GS, ref. 17-EURE-0014). K.B. acknowledges support from the Simons Collaboration on Extreme Wave Phenomena Based on Symmetries

%

\appendix

\onecolumngrid

\setcounter{figure}{0}
\setcounter{page}{1}
\renewcommand\thefigure{S\arabic{figure}} 
\renewcommand\thepage{S\arabic{page}}

\renewcommand{\thesubsection}{\Alph{subsection}}
\renewcommand{\thesubsubsection}{\thesubsection.\arabic{subsubsection}}
\makeatletter
\renewcommand{\p@subsection}{}
\renewcommand{\p@subsubsection}{}
\makeatother

\section*{Supplementary INFORMATION}
\subsection*{Transition Waves in Mechanical Metamaterials with Neighbor-Programmable Energy Landscapes}
\centerline{E.~Duval, G.~Risso, A. Zhang, V.~Tournat, and K.~Bertoldi}
\vspace{0.2in}

\subsection{Geometry}\label{sec:SIgeometry}
\begin{figure*}[!hpt]
\begin{center}
\includegraphics[width = 0.35\columnwidth]{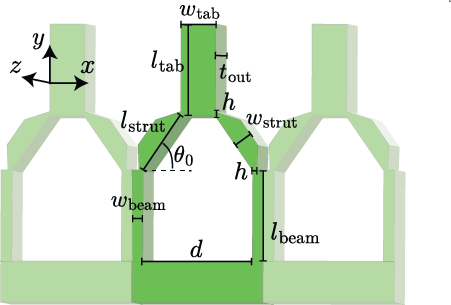} 
\caption{\label{fig:geom} Geometry of the considered structures.} 
\vspace{-15pt}
   \end{center}
\end{figure*}

All structures fabricated in this study consist of a one-dimensional (1D) array of unit cells sharing the geometry illustrated in Fig.~\ref{fig:geom}. Throughout the study, the following geometric parameters are kept constant: $t_\text{out} = 10$~mm, $w_\text{strut} = 3$~mm, $l_\text{strut} = 11$~mm, $w_\text{tab} = 6$~mm, $d = 18.4$~mm, and $l_\text{beam} = 15$~mm. Different combinations of vertical beam width ($w_\text{beam}$) and hinge thickness ($h$) are explored, as summarized in Table~\ref{tab:geom}.

\begin{table}[h!]
    \centering
    \begin{tabular}{l|c|c}
        \toprule
        \textbf{Structure} & \textbf{h [mm]} & \textbf{$w_\text{beam}$ [mm]} \\
        \midrule
        Structure \#a & 0.7 & 1.4 \\
        Structure \#b & 0.6 & 1.4 \\
        Structure \#c & 0.7 & 1.7 \\
        Structure \#d & 0.7 & 2.0 \\
        \bottomrule
    \end{tabular}
    \caption{Parameters defining the four structures considered in this study.}
    \label{tab:geom}
\end{table}

\newpage
\subsection{Fabrication}\label{sec:SIfabrication}
All samples investigated in this study are fabricated from nearly incompressible polyvinylsiloxane (PVS) elastomers (Elite Double 32, Zhermack, green color) using a casting approach. The molds are designed in the commercial CAD software \textit{Onshape} and fabricated by 3D printing with PLA filament (0.4 mm nozzle, Bambu Lab X1 Carbon, X1C). As part of this study, we fabricate two sets of samples: 1) samples with 9 units used for quasi-static testing and 2) samples with 32 units used to investigate wave propagation. 

\subsubsection{9-unit samples}
\begin{figure*}[!hpt]
\begin{center}
\includegraphics[width = 0.85\columnwidth]{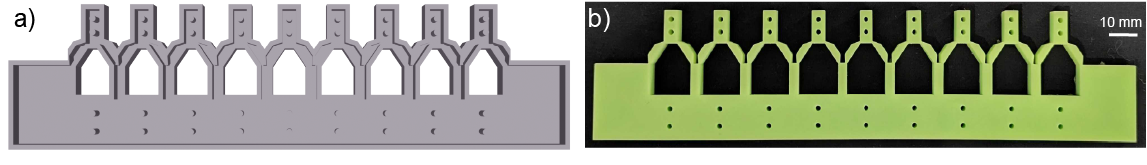} 
\caption{\label{fig:fabr9units} \textbf{9-unit sample} \textbf{a)} Mold used for sample casting. \textbf{b)} Example of a cured sample. } 
\vspace{-15pt}
   \end{center}
\end{figure*}

To fabricate these structures,  dedicated 9-unit molds were produced (Fig.~\ref{fig:fabr9units}). The choice of a 9-unit configuration provides a sufficient number of unit cells to capture representative behavior while keeping the overall sample size small enough to fit within the testing apparatus. The molds include precisely positioned pins with a diameter of $2.6$ mm that create small holes in the samples. These holes serve two purposes: (i) to enable attachment to both the moving and fixed translational stages, and (ii) to accommodate clip-on hold-down elements that constrain the height of selected unit cells during testing. 

The fabrication of a 9-unit sample was carried out as follows:
\begin{enumerate}
    \item A thin layer of Mann Release Mold 200 was applied to the mold surface.  
    \item The mold was filled with uncured PVS elastomer (mix ratio 1A:1B) and left to cure at room temperature (25$^\circ$C) for 30 minutes.  
    \item The cured part was carefully removed from the mold.  
    \item The sample was cured at room temperature for 15 days before being tested.  
\end{enumerate}

\subsubsection{32-unit samples}
\begin{figure*}[!hpt]
\begin{center}
\includegraphics[width = 0.85\columnwidth]{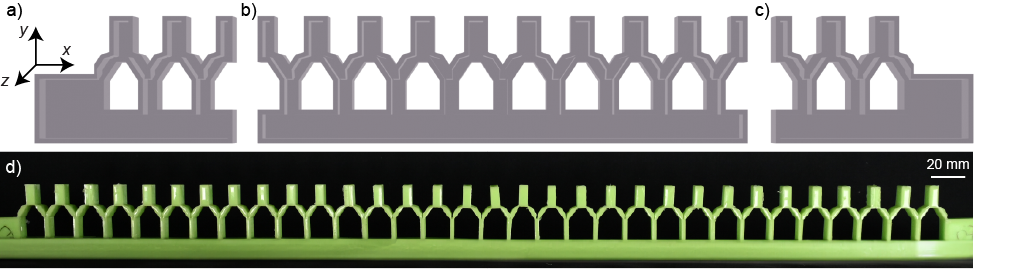} 
\caption{\label{fig:figure_molds} \textbf{32-unit samples}  \textbf{a)} Left-end mold. \textbf{b)} Middle mold. \textbf{c)} Right-end mold. \textbf{d)} Example of a casted 32-unit metamaterial. }
\vspace{-15pt}
   \end{center}
\end{figure*}

The 32-unit samples were manufactured in different parts that were subsequently bonded together. This modular design ensures both flexibility and scalability in the fabrication. Each part was obtained from one of three custom molds:  
\begin{itemize}
\item a left-end mold containing $2.5$ unit cells and a side block on the left (Fig.~\ref{fig:figure_molds}a),  
\item a middle mold comprising $8$ full unit cells and a half unit cell at each end(Fig.~\ref{fig:figure_molds}b),
\item a right-end mold containing $2.5$ unit cells and a side block on the right (Fig.~\ref{fig:figure_molds}c).  
\end{itemize}
A 32-unit sample consists of one left-hand part, three middle parts, and one right-hand part, with an example illustrated in Fig.~\ref{fig:figure_molds}. 

Each structure includes an integrated elastomeric foundation and lateral side blocks, cast as part of the same molds. The foundation layer extends approximately $10$~mm below the trusses, and the lateral side blocks are $25$~mm thick on each side. 

To manufacture the 32-unit sample, we proceed as follows:
\begin{enumerate}
    \item A thin layer of Mann Release Mold 200 is applied to the mold surface.  
    \item The molds are filled with uncured PVS elastomer (mix ratio 1A:1B) and left to cure at room temperature (25$^\circ$C) for 30 minutes.  
    \item The cured parts are carefully removed from the mold.  
    \item To assemble the 32-unit sample, one left-hand part, three middle parts, and one right-hand part are glued together by applying a thin layer of uncured PVS elastomer at the interfaces and pressing the parts together to ensure bonding.
    \item The assembled sample is cured at room temperature for 15 days before being tested. 
\end{enumerate}

\newpage
\subsection{Testing}\label{sec:SItest}
\subsubsection{Quasi-static tests}\label{sec:SIforcedisp}

\begin{figure*}[!hpt]
\begin{center}
\includegraphics[width = 0.99\columnwidth]{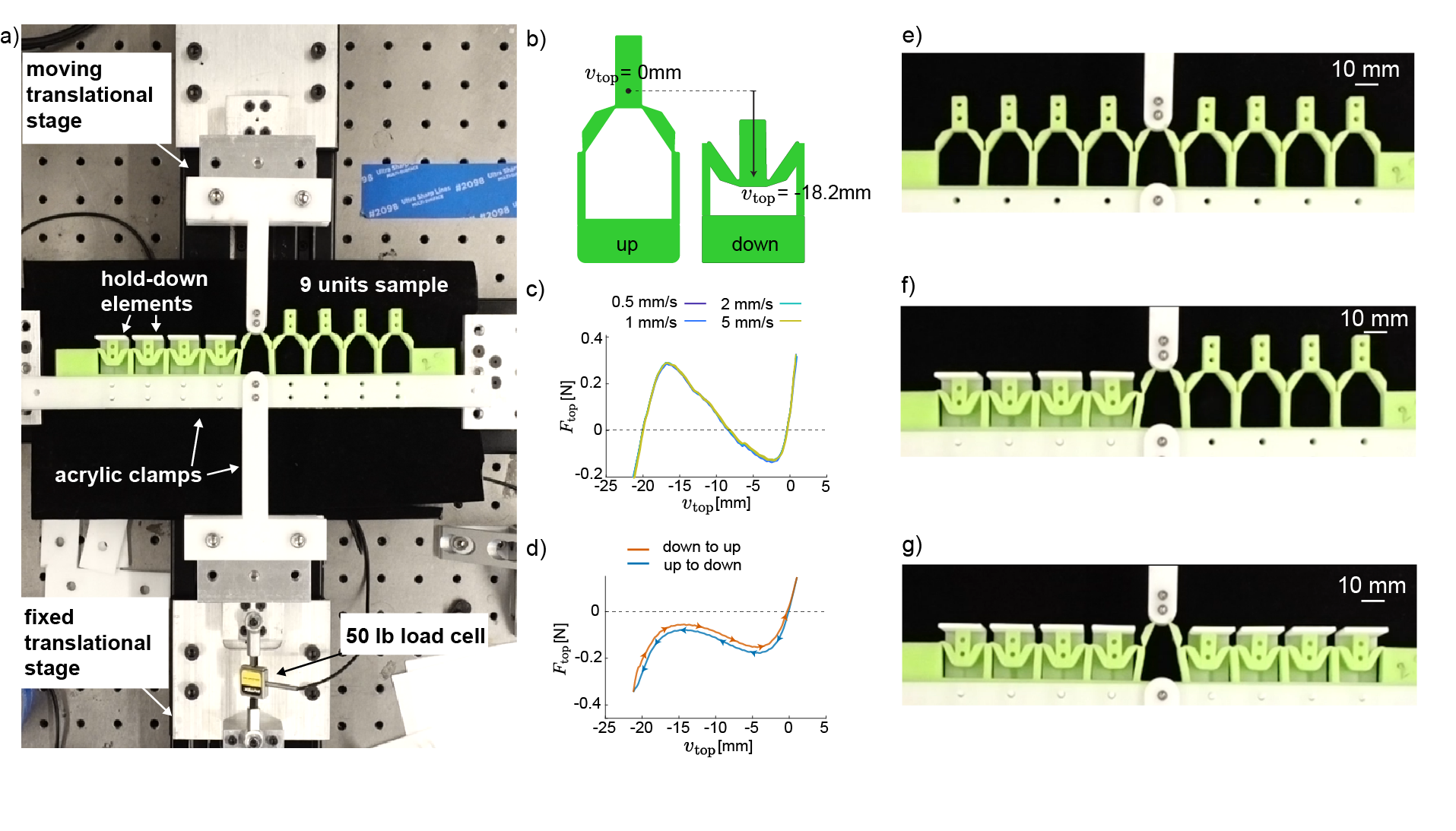} 
\caption{\label{fig:set_up_exp_quasi_statique} \textbf{Quasi-static test setup.} \textbf{a)} Experimental setup with a sample where the right side is up and the left side is down. The hold-down elements are used to fix boundary conditions and suppress unwanted motion during actuation. \textbf{b)} Schematics of the sign convention for $v_\text{top}$. \textbf{c)} Effect of the loading rate for the configuration with all units down. \textbf{d)} Full loading and unloading cycle for the all-up configuration. \textbf{e)} Case where all units surrounding the central one are in their up configuration. \textbf{f)} Case where the units to the left of the central one are kept in their down configuration, while those to the right remain up. \textbf{g)} Case where all units surrounding the central one are in their down configuration.} 
\vspace{-15pt}
   \end{center}
\end{figure*}

Quasi-static mechanical testing on the 9-unit samples is performed using two motorized translation stages (Thorlabs LTS300 and LTS150) and a $50$~lb capacity load cell (Futek LSB200). The base of the structure is rigidly attached to the load cell, while the pull tab of the central unit is connected to the translational stage via two screws, as shown in Fig.~\ref{fig:set_up_exp_quasi_statique}a. A vertical displacement $v_\text{top}$ is imposed at a constant speed of $2$~mm/s. In this study, we assume that $v_\text{top}=0$ mm corresponds to the up and stress-free configuration, and negative values of $v_\text{top}$ correspond to the closing of the structure (Fig.~\ref{fig:set_up_exp_quasi_statique}b). To ensure that the response is quasi-static, we performed experiments at loading rates of $0.5$~mm/s, $1$~mm/s, and $5$~mm/s.  Fig.~\ref{fig:set_up_exp_quasi_statique}c shows negligible differences in the response for a structure \#a, in the configuration with all units down.

\begin{figure*}[!hpt]
\begin{center}
\includegraphics[width = 0.99\columnwidth]{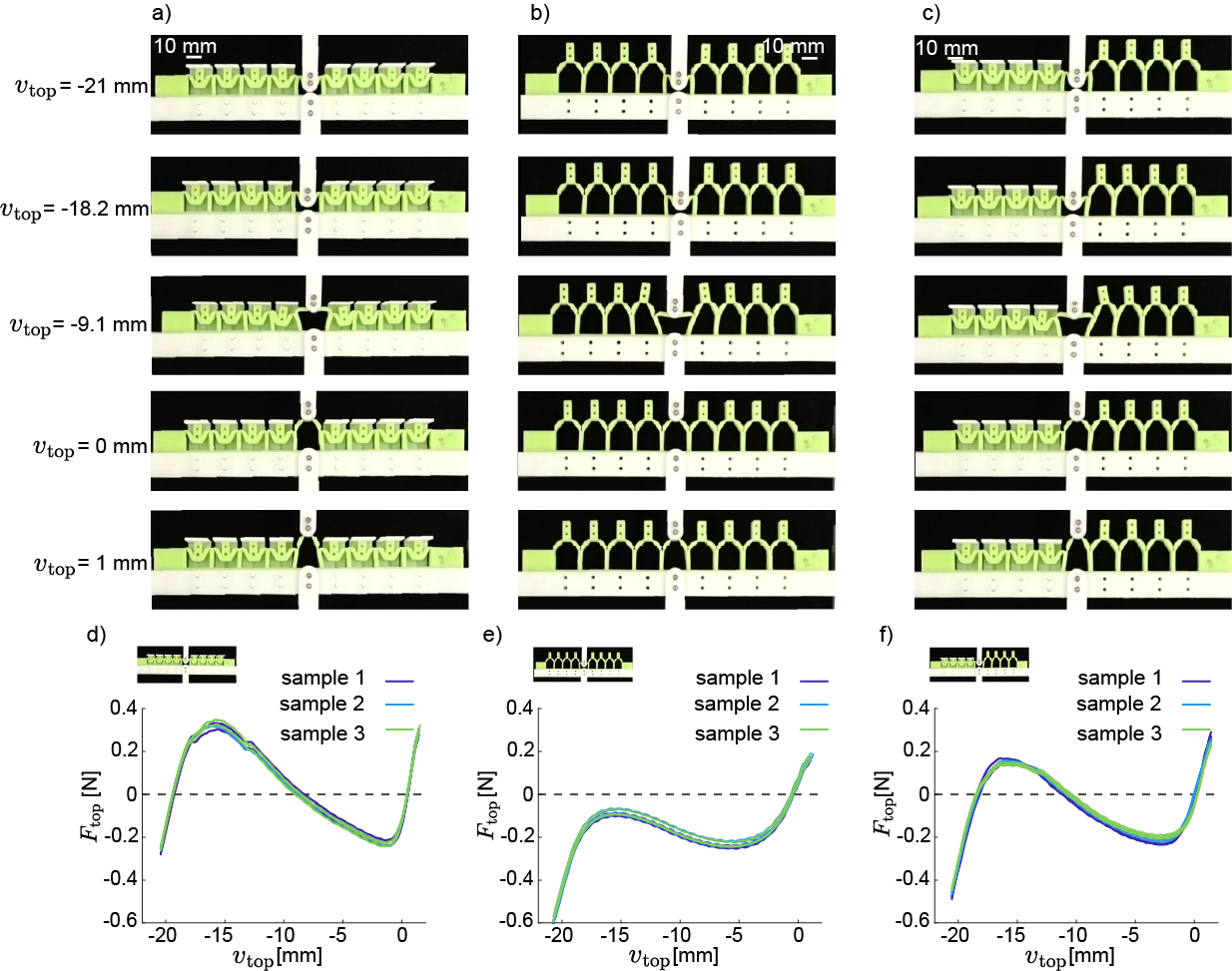} 
\caption{\label{fig:exp_quasi_statique} \textbf{a-c)} Snapshots of quasi-static deformation at successive imposed displacement $v_\text{top}$ steps. All units down (a), all units up (a), left side is down, and right side is up (c). \textbf{d-f)} Experimental force-displacement curves for structure \#a ($h = 0.7$~mm, $w_\text{beam} = 1.4$~mm) for all units down (d). all units up (e), one side up, one side down (f).} 
\vspace{-15pt}
   \end{center}
\end{figure*}

In this study, we focus on the pop-up behavior. Consequently, although the tested unit is first lowered and then raised, our analysis considers only the unloading phase. Furthermore, each sample is tested in three different configurations:
\begin{itemize}
\item a test in which all units surrounding the central one are in their up configuration (Fig.~\ref{fig:set_up_exp_quasi_statique}e and Fig.~\ref{fig:exp_quasi_statique}b);
\item a test in which the units to the left of the central one are kept in their down configuration, while those to the right remain up (Fig.~\ref{fig:set_up_exp_quasi_statique}f and Fig.~\ref{fig:exp_quasi_statique}c);
\item a test in which all units surrounding the central one are in their down configuration (Fig.~\ref{fig:set_up_exp_quasi_statique}g and Fig.~\ref{fig:exp_quasi_statique}a).
\end{itemize}
To maintain a unit in its down state, we employ 3D-printed hold-down elements (see Figs.~\ref{fig:set_up_exp_quasi_statique}a and Fig.~\ref{fig:exp_quasi_statique}a-c). When these elements are installed, the corresponding units cannot move upward beyond the hold-down elements but can still undergo larger downward displacements beneath  (i.e. $v_\text{top} < -18.2$ mm).

Finally, for completeness, in  Fig.~\ref{fig:set_up_exp_quasi_statique}d, we show a full loading cycle for Structure \#a with all units up, including both lowering and raising phases. Here, the test begins from a preloaded position of $v_\text{top} = 1$~mm, followed by a downward translation to $v_\text{top} = - 21$~mm, a few millimeters beyond the full truss closure. The recorded force–displacement curves nearly overlap, indicating minimal hysteresis in the system.

\begin{figure*}[!hpt]
\begin{center}
\includegraphics[width = 0.99\columnwidth]{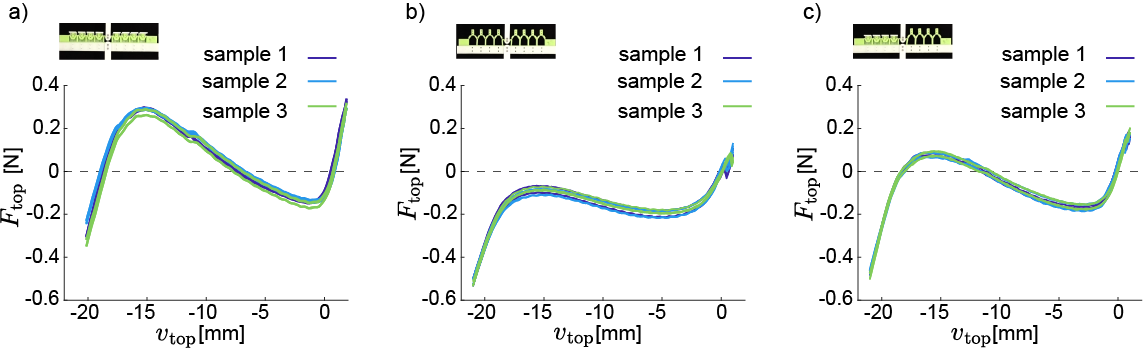} 
\caption{\label{fig:f_u_exp_down_06} Experimental force-displacement curves for for structure \#b  ($h = 0.6$~mm, $w_\text{beam} = 1.4$~mm) \textbf{a)} All units down \textbf{b)} All units up \textbf{c)} One side up, one side down. } 
\vspace{-15pt}
   \end{center}
\end{figure*}

To ensure repeatability, we fabricate and test three identical 9-unit samples for Structures \#a and \#b. For each sample, we perform two loading-unloading cycles and repeat twice, resulting in four curves per sample and twelve curves per geometry in total. The resulting force-displacement curves are shown in Fig.~\ref{fig:exp_quasi_statique} for Structure $\#$a (with $h = 0.7$~mm and $w_\text{beam} = 1.4$~mm) and in Fig.~\ref{fig:f_u_exp_down_06} for Structure $\#$b (with $h = 0.6$~mm, $w_\text{beam} = 1.4$~mm). The results exhibit high repeatability with minimal variations across cycles and casts. 

\begin{figure*}[!hpt]
\begin{center}
\includegraphics[width = 0.99\columnwidth]{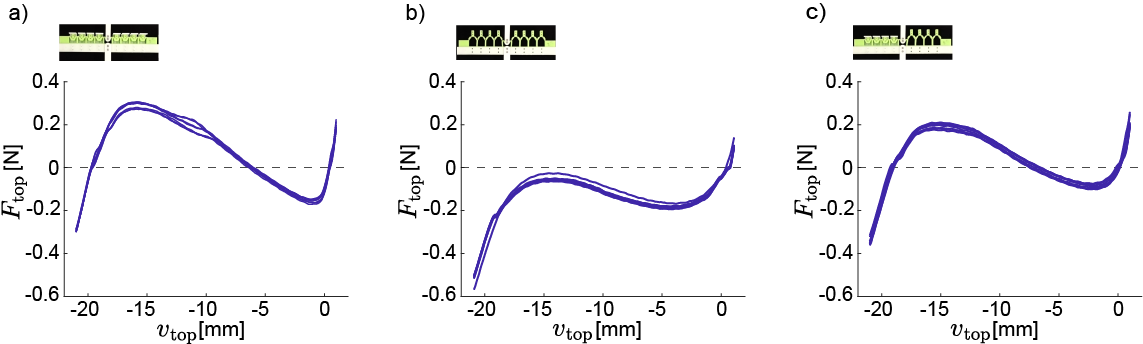} 
\caption{\label{fig:f_u_exp_down_c} Experimental force-displacement curves for for structure \#c  ($h = 0.7$~mm, $w_\text{beam} = 1.7$~mm) \textbf{a)} All units down \textbf{b)} All units up \textbf{c)} One side up, one side down. } 
\vspace{-15pt}
   \end{center}
\end{figure*}

\begin{figure*}[!hpt]
\begin{center}
\includegraphics[width = 0.99\columnwidth]{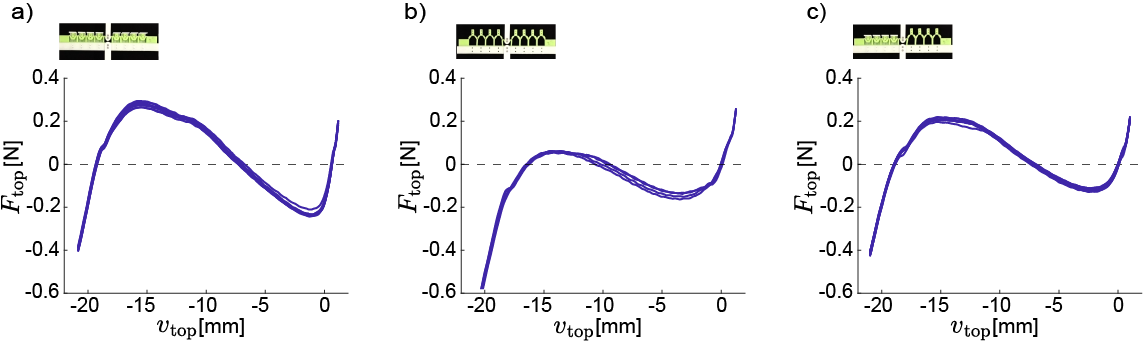} 
\caption{\label{fig:f_u_exp_down_d} Experimental force-displacement curves for for structure \#d ($h = 0.7$~mm, $w_\text{beam} = 2.0$~mm) \textbf{a)} All units down \textbf{b)} All units up \textbf{c)} One side up, one side down. } 
\vspace{-15pt}
   \end{center}
\end{figure*}

In addition, we fabricated and tested one 9-unit sample for each of structures \#c and \#d, and the corresponding results are presented in Figs.~\ref{fig:f_u_exp_down_c} and~\ref{fig:f_u_exp_down_d} respectively.

\begin{figure}[h!]
    \centering
    \includegraphics[width=0.7\linewidth]{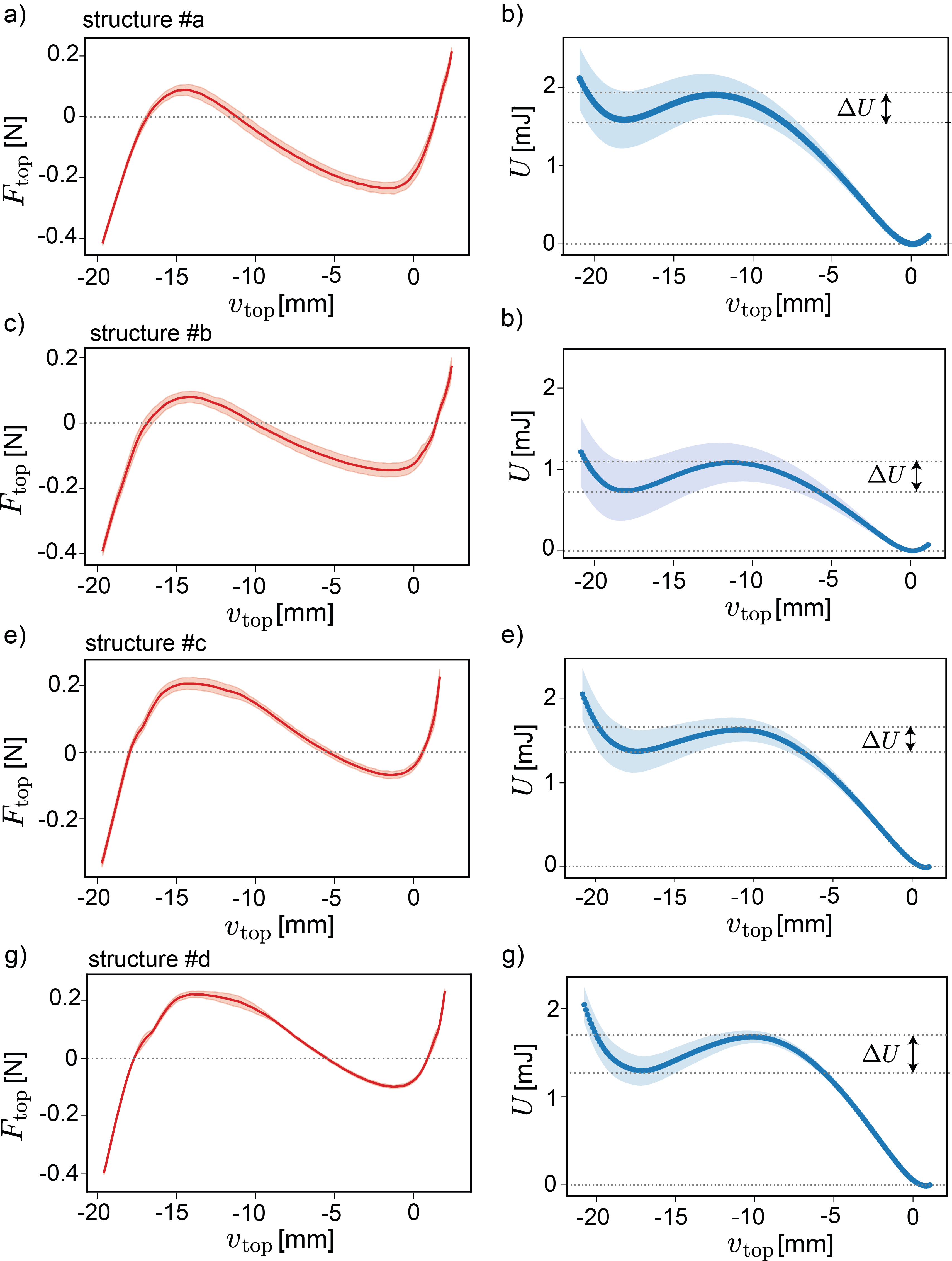}
    \caption{Experimental force–displacement responses (left column) and corresponding reconstructed energy landscapes (right column). From top to bottom, rows \textbf{(a–b)}, \textbf{(c–d)}, \textbf{(e–f)}, and \textbf{(g–h)} correspond to Structures~\#a, \#b, \#c, and \#d, respectively. Shaded areas indicate experimental uncertainty.}
    \label{fig:figure_exp_energi}
\end{figure}

From the experimental quasi-static force–displacement curve $F_{\text{top}}$–$v_{\text{top}}$, we compute the potential energy $U$ of the central unit in a 9-cell metamaterial as a function of $v_{\text{top}}$ (Fig.~\ref{fig:figure_exp_energi}). Specifically, the potential energy is obtained by numerically integrating the measured force:
\begin{equation}
    U(v_{\text{top}}) = \int F_{\text{top}}(v)\,\mathrm{d}v,
\end{equation}
with the reference chosen such that $U(v_{\text{top}}=0)=0$.

We then define $\Delta U$ as the energy barrier associated with the transition from the down state to the up state when the von Mises trusses on one side of the central unit are held in the down state and those on the opposite side in the up state (see Fig.~\ref{fig:figure_exp_energi}, right column). Fig.~\ref{fig:figure_exp_energi} presents the experimental force–displacement curves and corresponding energy landscapes for structures \#a–d in the one-side-up, one-side-down configuration, while Table~\ref{tab:geom_speed} reports the associated values of $\Delta U$.

\begin{table}[h!]
    \centering
    \begin{tabular}{l|c|c|c}
        \toprule
        \textbf{Structure}  & \textbf{ Experimental $c_{\text{wave}}$ [units/s] } & \textbf{$\Delta U$ [mJ]} \\ 
        \midrule
        Structure \#a & 54.1 $\pm$ 3.2 & 0.31\\
        Structure \#b & 40.1 $\pm$ 3.9  & 0.34 \\
        Structure \#c & 37.2 $\pm$ 4.1  & 0.37 \\
        Structure \#d & No propagation  & 0.42 \\
        \bottomrule
    \end{tabular}
    \caption{Summary of the speed $c_{wave}$ and corresponding $ U_\text{down}$ and  $\Delta U$.}
    \label{tab:geom_speed}
\end{table}

\subsubsection{Dynamic tests}
\begin{figure*}[!hpt]
\begin{center}
\includegraphics[width = 0.6\columnwidth]{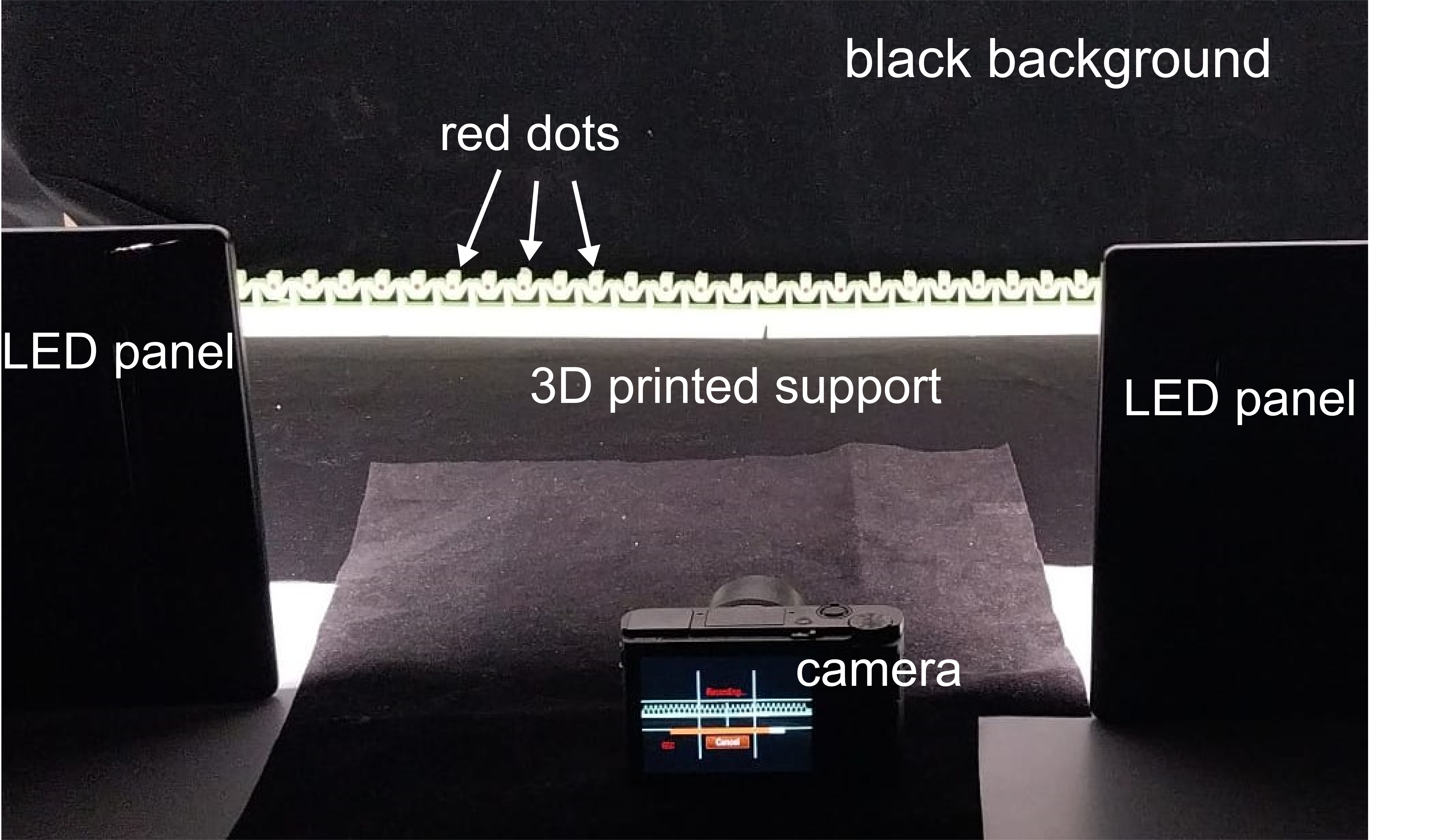} 
\caption{\label{fig:exp_set_up} Overview of the dynamical test set-up of a structure at zero curvature.  }
\vspace{-15pt}
   \end{center}
\end{figure*}

Dynamic experiments are performed on 32-unit structures. Each unit’s top pull tab is marked with a red dot to enable position tracking during testing. The base of the structure is clamped into a 3D-printed PLA support to ensure mechanical stability throughout the dynamic experiments. The setup is positioned against a black background and illuminated with two NEEWER NL-192AI LED panels (see Fig.~\ref{fig:exp_set_up}). All pull tabs are initially set manually to the down state, corresponding to $v_\text{top}^j = - 18.2$~mm for $j = 1, \ldots, 32$. Then, the $i$-th unit is manually lifted to the stress-free up configuration ($v_\text{top}^i = 0$~mm) and released. Depending on the structure, this localized perturbation may trigger a propagating transition wave along the chain. The experiments are recorded using a Sony RX100V camera at 480 fps. Vertical displacements of the top masses are extracted through frame-by-frame tracking of the red markers using a custom MATLAB script.

\clearpage
\newpage
\subsection{Numerical Model}\label{sec:SImodel}

To capture the mechanical behavior of the considered structures, we developed a minimal mass-spring model reflecting the primary deformation mechanisms observed in the experiments. 

\subsubsection{Unit cell description}

\begin{figure*}[!hpt]
\begin{center}
\includegraphics[width = 0.99\columnwidth]{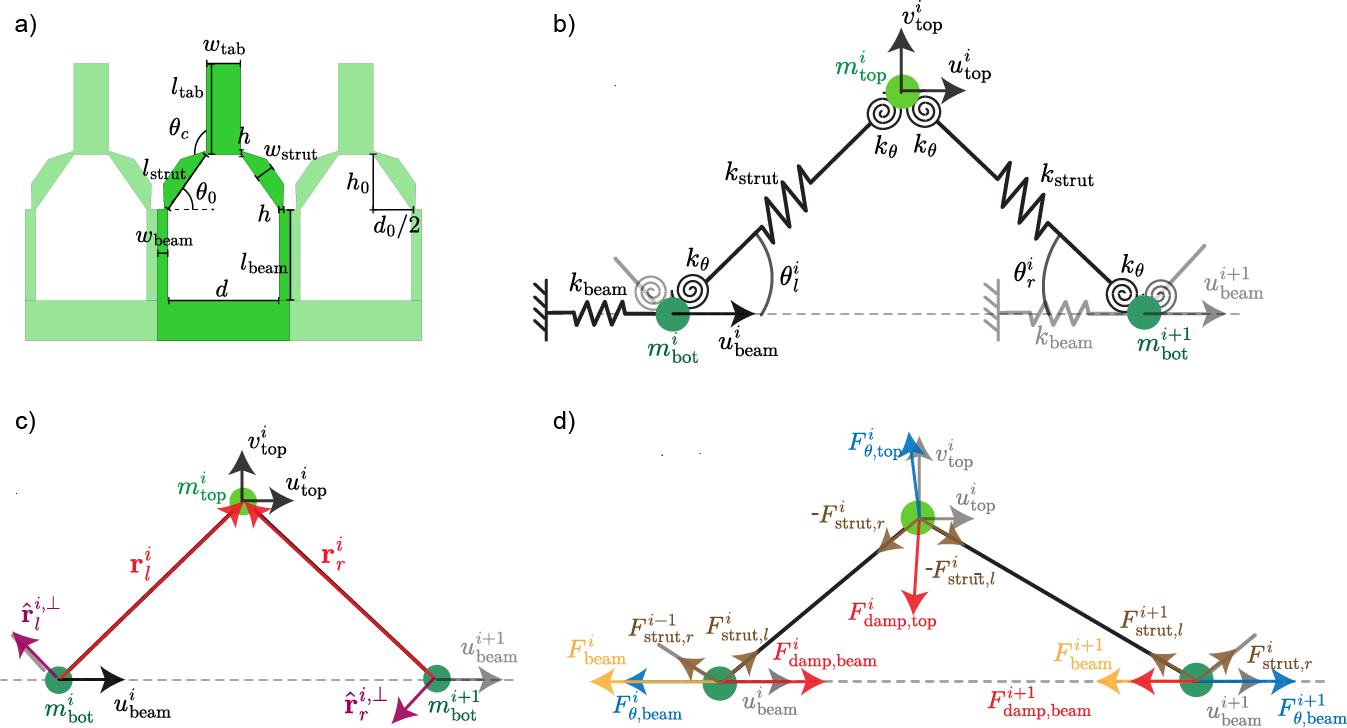} 
\caption{\label{fig:sketches} Geometry and modeling of the unit cells. \textbf{a)} Geometric layout of a unit cell. \textbf{b)} Schematic representation of the mechanical model used to describe the unit cell's behavior. \textbf{c)} Representation of the vectors used in the analytical description of the unit cell’s behavior. \textbf{d)} Forces acting on the masses: axial–strut forces (brown), lateral (ground) spring forces (orange), viscous damping (red), and hinge-induced (torsional) forces (blue).}
\vspace{-15pt}
   \end{center}
\end{figure*}

As shown in Fig.~\ref{fig:sketches}, each unit cell is modeled as two concentrated masses connected by a network of linear and torsional springs that reproduce the geometry of the structure. Specifically,  the $i$-th unit cell consists of two concentrated masses ($m^i_\text{bot}$ with a single degree of freedom and $m^i_\text{top}$ with two degrees of freedom), three axial linear springs, and four torsional linear springs. If $m^i_\text{bot}$ is located at ($x_i$, 0) in the initial configuration, then $m^{i+1}_\text{bot}$ is located at ($x_i+ d_0$, 0) and $m^{i}_\text{top}$ is located at ($x_i+d_0/2$, $h_0$), where  $d_0 = d - w_\text{tab}$  and $h_0 = \sqrt{l_\text{strut}^2 - (d_0/2)^2}$.

\smallskip

\paragraph*{Concentrated masses.} The $i$-th unit cell contains two lumped masses, $m^i_{\text{top}}$ and $m^i_{\text{bot}}$. 
\begin{itemize}
    \item $m^i_{\text{top}}$ has two degrees of freedom: horizontal displacement $u^i_{\text{top}}$ 
and vertical displacement $v^i_{\text{top}}$. It represents the mass of the vertical tab 
(of length $l_\text{tab}$ and width $w_\text{tab}$) located at the top of each unit cell, as well as half of the mass 
of each of the two adjacent inclined struts, so that
    \begin{equation}
    m^i_\text{top}= w_\text{tab}\, l_\text{tab }\, t_\text{out}\,\rho + m_\text{strut},
    \end{equation}
where $\rho$ is the density of the material and $m_\text{strut}$ is the mass of an inclined strut.
\item $m^i_{\text{bot}}$ has a single degree of freedom (horizontal 
displacement $u^i_{\text{beam}}$) and represents the mass of the vertical beam together with the remaining half of the mass of the adjacent inclined strut. The vertical beam is modeled as a cantilever, with its inertia represented by an equivalent lumped mass of 0.236 $\rho\,2w_\text{beam}\,l_\text{beam}\,t_\text{out}$
~\cite{BlevinsFormulas}. By adding half the mass of the adjacent inclined struts, the total effective mass becomes
    \begin{equation}
    m^i_{beam}= 0.236*2 w_\text{beam}\,l_\text{beam}\,t_\text{out}+  m_\text{strut}.
    \end{equation}
\end{itemize}

To simplify the formulation of the system’s equations of motion, we define, for the $i$-th unit cell, two vectors $\mathbf{r}_l^i$ and $\mathbf{r}_r^i$, which connect $m^i_{\text{bot}}$ and $m^{i+1}_{\text{bot}}$ to $m^i_{\text{top}}$, respectively (see Fig.~\ref{fig:sketches}b)
\begin{subequations}
\begin{equation}
    \mathbf{r}_l^i = \begin{pmatrix} u_\text{top}^i - u_\text{beam}^i + d_0/2 \\ h_0 + v_\text{top}^i \end{pmatrix} ,
\end{equation}

\begin{equation}
    \mathbf{r}_r^i = \begin{pmatrix} u_\text{top}^i - u_\text{beam}^{i+1} - d_0/2 \\ h_0 + v_\text{top}^i \end{pmatrix} .
\end{equation}
\end{subequations}

\paragraph*{Springs.}
The $i$-th unit cell contains two axial linear springs of stiffness $k_\text{strut}$, one axial linear spring of stiffness $k_\text{beam}$, and four linear rotational springs of stiffness $k_\theta$. 
\begin{itemize}
    \item The axial springs with stiffness 
$k_\text{strut}$ represents the potential extension and compression of the inclined struts under loading. The stiffness $k_\text{strut}$ is defined as
    \begin{equation}
    \label{eq:kl}
        k_\text{strut} = \frac{E t_\text{out} w_\text{strut}}{l_\text{strut}}=3.5~\frac{kN}{m} .
    \end{equation}

    These springs generate a force 
    \begin{equation}
    \label{eq:ax_top}
        \mathbf{F}_{\text{strut},l}^{i} =  k_\text{strut} \left(l_l^i - l \right) \frac{\mathbf{r}_l^i}{l_l^i}\quad\quad  \text{and} \quad\quad \mathbf{F}_{\text{strut},r}^{i} = k_\text{strut} \left(l_r^i - l \right) \frac{\mathbf{r}_r^i}{l_r^i} ,
    \end{equation}
    where
    \begin{equation}
    l_{l}^i = \|\mathbf{r}_{l}^i \| \quad\quad  \text{and} \quad\quad  l_{r}^i = \|\mathbf{r}_{r}^i \| .
\end{equation}

 \item The axial springs with stiffness $k_\text{beam}$ represent the transverse deformation of the vertical beams under lateral loading. 
The stiffness $k_\text{beam}$ is defined as~\cite{Timoshenko1962}:
\begin{equation}
\label{eq:kg}
    k_\text{beam} = \frac{3EI}{l_\text{beam}^3}, \quad \text{with} \quad I = \frac{(2w_\text{beam})^3 t_\text{out}}{12} .
\end{equation}
For structures \#a and \#b, $k_\text{beam} = 21.1~\mathrm{N/m}$, for structure \#c, $k_\text{beam} = 37.8~\mathrm{N/m}$, and for structure \#d, $k_\text{beam} = 61.7~\mathrm{N/m}$.
These springs generates an horizontal force acting on $m^i_{\text{bot}}$ equal to
    \begin{equation}
    \label{eq:x2_beam}
       \mathbf{F}_{\text{beam}}^i = - k_\text{beam} u_\text{beam}\mathbf{e}_x.
    \end{equation}
 where $\mathbf{e}_x$ is the unit vector along the x direction   
\item The torsional springs with stiffness $k_\theta$ model the response of the thin ligaments located at the junctions between the struts and the pull bars or vertical base beams. The springs connecting the inclined struts to the vertical beams and to the pull tab have stress-free configurations at angles $\theta_0$ and $\pi/2 - \theta_0$, respectively, where $\theta_0 = 55.7^{\circ}$ denotes the angle that the inclined struts form with the horizontal direction in the undeformed configuration. The stiffness $k_\theta$ is determined by calibrating the model against quasi-static experimental data. For the geometries considered in this study, $k_\theta = 0.77~\mathrm{mN{\cdot}m/rad}$ for structure \#b with $h = 0.6~\mathrm{mm}$, and $k_\theta = 1.02~\mathrm{mN{\cdot}m/rad}$ for all structures with $h = 0.7~\mathrm{mm}$ (i.e., structures \#a, \#c, and \#d). Details of the calibration procedure are provided in Section~\ref{sec:calibration}.

Each torsional spring exerts a restoring torque proportional to the angular deviation from its stress-free configuration. It follows that the torques generated on the left side at the top hinge, $\tau_{lt}^i$, on the left side at the bottom hinge, $\tau_{lb}^i$, on the right side at the top hinge, $\tau_{rt}^i$, on the right side at the bottom hinge, $\tau_{rb}^i$,  are given by

\begin{subequations}
    \begin{equation}
    \tau_{lb}^i = -k_\theta\left(\theta_l^i - \theta_{0} \right) ,
    \end{equation}
    \begin{equation}
    \tau_{lt}^i = -k_\theta\left[\left(\frac{\pi}{2}-\theta_l^i\right) - \left(\frac{\pi}{2}-\theta_{0}\right) \right]=-k_\theta\left(\theta_l^i - \theta_{0} \right)=\tau_{lb}^i ,
    \end{equation}
    \begin{equation}
    \tau_{rb}^i = -k_\theta\left(\theta_r^i - \theta_{0} \right),
    \end{equation}
    \begin{equation}
    \tau_{rt}^i = -k_\theta\left[\left(\frac{\pi}{2}-\theta_r^i\right) - \left(\frac{\pi}{2}-\theta_{0}\right) \right]=-k_\theta\left(\theta_r^i - \theta_{0} \right)=\tau_{rb}^i ,
    \end{equation}
\end{subequations}
where
\begin{equation}
    \theta_{l}^i = \arctan{\left(\frac{\mathbf{r}_{l}^i \cdot \mathbf{e}_y}{\mathbf{r}_{l}^i \cdot \mathbf{e}_x}\right)},\quad\quad\quad \theta_{r}^i = \pi - \arctan{\left(\frac{\mathbf{r}_{r}^i \cdot \mathbf{e}_y}{\mathbf{r}_{r}^i \cdot \mathbf{e}_x}\right)}.
\end{equation}
To account for potential contact between the inclined struts and the vertical tab at large rotations, the top hinges are assigned an effective nonlinear rotational stiffness $\tilde{k}_\theta$, which may differ from 
$k_\theta$ depending on the deviation angles. More specifically, since in the experiments the upper ends of the inclined struts come into contact with the vertical tabs when the angular deviation of the top hinges exceeds a critical value $\theta_c= 108^{\circ}$ (see Fig.~\ref{fig:sketches}a), the  effective nonlinear rotational stiffness for the top hinges is  defined as~\cite{Zareei2020}: 

\begin{equation}
\label{eq:k_theta}
\tilde{k}_\theta = \left\{
    \begin{array}{ll}
        k_\theta & \quad \mathrm{ if } \quad \theta^i-\theta_0 < \theta_{c} \\
        1000 k_\theta \left(\theta^i -\theta_0- \theta_c \right)^2 & \quad \mathrm{ if } \quad \theta^i-\theta_0 > \theta_{c} 
    \end{array}
\right.
\end{equation}

It follows from Eq.~(\ref{eq:k_theta}) that when there is contact (i.e. for $\theta^i-\theta_0>\theta_c$), $\tilde{k}_\theta \gg k_\theta$.

\end{itemize}

\paragraph*{Dissipation.} Despite the complex dissipative behavior of elastomers, all simulations in this study employ a linear viscous damping model, so that : 
 
    \begin{equation}
    \label{eq:damp_top}
        \mathbf{F}_{\text{damp},\text{top}}^{i} = - c_\text{top}
    \begin{pmatrix}
            \dot{u}_\text{top}^i \\ \dot{v}_\text{top}^i
    \end{pmatrix}
    \end{equation}
    and
    \begin{equation}
    \label{eq:damp_beam}
        F_{\text{damp},\text{beam}}^{i} = - c_\text{beam} \dot{u}_\text{beam}^i 
    \end{equation}
where $c_\text{beam}$ and $c_\text{top}$ are the damping coefficients associated with the top and beam masses, respectively. 
The values of these coefficients were determined by matching the computational and experimental wave propagation speeds in structure \#a, and were subsequently kept constant for all four structures analyzed in this study: $c_\text{top} = 0.016~\mathrm{N\cdot s/m}$ and $c_\text{beam} = 0.022~\mathrm{N\cdot s/m}$.

\subsubsection{Equations of motion}

We now consider a chain composed of $N$ unit cells. The leftmost and rightmost beam masses ($m^0_\text{bot}$ and $m^{N+1}_\text{bot}$) are fixed (i.e. $u^0_\text{beam} =0$ and $u^{N+1}_\text{beam} =0$). We  determine the equation of motion for the system as:

\begin{subequations}
\label{eq:newton_laws}
\begin{align}
m_\text{top} \ddot{u}_\text{top}^i &= - \mathbf{F}^i_{{\text{strut}},l}\cdot\mathbf{e}_x - \mathbf{F}^i_{{\text{strut}},r}\cdot\mathbf{e}_x+ \mathbf{F}^{i}_{\theta, \text{top}}\cdot\mathbf{e}_x + \mathbf{F}^{i}_{\text{damp}, \text{top}}\cdot\mathbf{e}_x 
\\[6pt]
m_\text{top}\ddot{v}_\text{top}^i &= - \mathbf{F}^i_{{\text{strut}},l}\cdot\mathbf{e}_y - \mathbf{F}^i_{{\text{strut}},r}\cdot\mathbf{e}_y + \mathbf{F}^{i}_{\theta, \text{top}}\cdot\mathbf{e}_y + \mathbf{F}^{i}_{\text{damp}, \text{top}}\cdot\mathbf{e}_y
\\[6pt]
m_\text{bot}\ddot{u}_\text{beam}^i &= \mathbf{F}^i_{{\text{strut}},l}\cdot\mathbf{e}_x + \mathbf{F}^{i-1}_{{\text{strut}},r}\cdot\mathbf{e}_x + \mathbf{F}^i_{\text{beam}}\cdot\mathbf{e}_x + F^i_{\theta, \text{beam}} + F^i_{\text{damp}, \text{beam}}
\end{align}
\end{subequations}
where $\mathbf{F}_{\theta, \text{top}}^{i}$ and $F_{\theta, \text{beam}}^{i}$ denote the forces generated on the two masses by the  torques.  They are obtained through the moment–arm relationship as
    \begin{subequations}
    \label{eq:theta_top_beam}
    \begin{equation}
    \label{eq:theta_top}
    \mathbf{F}_{\theta, \text{top}}^{i} = \tau^i_{lb}  \frac{\hat{\mathbf{r}}_l^{i, \perp}}{l_l^i} + \tau^i_{rb}  \frac{\hat{\mathbf{r}}_r^{i, \perp}}{l_r^i}= -k_\theta \left[\left(\theta_l^i - \theta_{0} \right) \frac{\hat{\mathbf{r}}_l^{i, \perp}}{l_l^i} + \left(\theta_r^i -  \theta_{0}\right) \frac{\hat{\mathbf{r}}_r^{i, \perp}}{l_r^i} \right]
    \end{equation}
    \begin{equation}
    \label{eq:theta_beam}
    F_{\theta, \text{beam}}^{i} = \left(-\tau^i_{lt}  \frac{\hat{\mathbf{r}}_l^{i, \perp}}{l_l^i} - \tau^{i-1}_{rt}  \frac{\hat{\mathbf{r}}_r^{i-1, \perp}}{l_r^{i-1}}\right) \cdot \mathbf{e}_x =\tilde{k}_\theta\left[\left(\theta_l^i - \theta_{0} \right) \frac{\mathbf{r}_l^{i} \cdot \mathbf{e}_y}{(l_l^i)^2} + \left(\theta_r^{i-1} - \theta_{0} \right) \frac{\mathbf{r}_r^{i-1} \cdot \mathbf{e}_y }{(l_r^{i-1})^2} \right]    \end{equation}
    \end{subequations}
where
$\hat{\mathbf{r}}_l^{i,\perp} = (-\mathbf{r}_l^{i} \cdot \mathbf{e}_y, \mathbf{r}_l^{i} \cdot \mathbf{e}_x)/l_l^i, \quad \hat{\mathbf{r}}_r^{i,\perp} = (- \mathbf{r}_r^{i} \cdot \mathbf{e}_y, \mathbf{r}_r^{i} \cdot \mathbf{e}_x) /l_r^i$ denote unit vectors perpendicular to $\mathbf{r}_l^i$ and $\mathbf{r}_r^i$, respectively,
each obtained by a $+90^{\circ}$ (counterclockwise) rotation.

Substitution of Eqns. (\ref{eq:ax_top}), (\ref{eq:x2_beam}),  (\ref{eq:theta_top_beam}),  (\ref{eq:damp_top}), and (\ref{eq:damp_beam}) into Eq.~(\ref{eq:newton_laws}) yields
 
\begin{subequations}
\label{eq:newton_laws1}
\begin{align}
m_\text{top} \ddot{u}_\text{top}^i &= 
-k_\text{strut}(l_l^i-l)\frac{r_{l}^i \cdot \mathbf{e}_x}{l_l^i}
-k_\text{strut}(l_r^i-l)\frac{\mathbf{r}_{r}^i \cdot \mathbf{e}_x}{l_r^i}
+k_\theta \left[
(\theta_l^i-\theta_{l,0})\frac{\mathbf{r}_{l}^i \cdot \mathbf{e}_y}{(l_l^i)^2}
+(\theta_r^i-\theta_{r,0})\frac{\mathbf{r}_{r}^i \cdot \mathbf{e}_y}{(l_r^i)^2}
\right]
-c_\text{top}\dot{u}_\text{top}^i, 
\\[6pt]
m_\text{top}\ddot{v}_\text{top}^i &=
-k_\text{strut}(l_l^i-l)\frac{\mathbf{r}_{l}^i \cdot \mathbf{e}_y}{l_l^i}
-k_\text{strut}(l_r^i-l)\frac{\mathbf{r}_{r}^i \cdot \mathbf{e}_y}{l_r^i}
-k_\theta\left[
(\theta_l^i-\theta_{l,0})\frac{\mathbf{r}_{l}^i \cdot \mathbf{e}_x}{(l_l^i)^2}
+(\theta_r^i-\theta_{r,0})\frac{\mathbf{r}_{r}^i \cdot \mathbf{e}_x}{(l_r^i)^2}
\right]
- c_\text{top} \dot{v}_\text{top}^i, 
\\[6pt]
m_\text{bot}\ddot{u}_\text{beam}^i &= 
-k_\text{beam}\,u_\text{beam}^i 
-c_\text{beam}\dot{u}_\text{beam}^i 
+k_\text{strut}(l_l^i-l)\frac{\mathrm{r}_{l}^i \cdot \mathbf{e}_x}{l_l^i}
+k_\text{strut}(l_r^{i-1}-l)\frac{\mathbf{r}_{r}^{i-1} \cdot \mathbf{e}_x}{l_r^{i-1}} \notag \\[-2pt]
&\quad
+ \tilde{k}_\theta\left[\left(\theta_l^i - \theta_{0} \right) \frac{\mathbf{r}_l^{i} \cdot \mathbf{e}_y}{(l_l^i)^2} + \left(\theta_r^{i-1} - \theta_{0} \right) \frac{\mathbf{r}_r^{i-1} \cdot \mathbf{e}_y }{(l_r^{i-1})^2} \right].
\end{align}
\end{subequations}
 that are integrated numerically over
$t \in [0,1]~\mathrm{s}$ with a nominal time step 
$\Delta t = 1/480~\mathrm{s}$ using the \texttt{odeint} solver from \textsc{JAX} (which employs an adaptive step-size Runge-Kutta method).

\subsubsection{Quasi-static simulations}\label{sec:calibration} 

Quasi-static numerical force–displacement simulations are employed both to calibrate the torsional stiffness $k_{\theta}$ of the model and to characterize the effective energy barrier governing wave propagation. Although the full equations of motion are integrated using a second-order dynamic solver, the imposed displacement varies sufficiently slowly that inertial effects are negligible. The resulting response therefore corresponds to a sequence of static equilibria under incremental loading.

As in the experiments described in Section~\ref{sec:SIforcedisp}, we consider a structure composed of $9$ units and impose a displacement on the central unit ($i = 4$). The prescribed vertical displacement $v_{\mathrm{target}}$ enforces a smooth transition from the lower to the upper equilibrium position over a total duration $T_{\mathrm{total}} = 10$~s:
\begin{equation}
\label{eq:y_target}
v_{\mathrm{target}}(t) = 2 h_0 \left(\frac{t}{T_\text{total}} - 1\right), 
\quad 
t \in \left[0, T_{\mathrm{total}}\right].
\end{equation}

This displacement control is implemented via a virtual control force with stiffness $k_{\mathrm{control}}$, chosen to be much larger than the intrinsic stiffness of the system:
\begin{equation}
\label{eq:force_control}
F_{\mathrm{control}} = - k_{\mathrm{control}} 
\left(v_\text{top}^i(t) - v_{\mathrm{target}}(t)\right).
\end{equation}
The control force acts vertically on the selected top mass and enters the equations of motion~(\ref{eq:newton_laws}) as an external force term.

The different quasi-static configurations are obtained by selectively constraining the motion and prescribing the initial state of the side units. Units that are left free evolve according to the full equations of motion, with both vertical and horizontal degrees of freedom active; unless otherwise specified, they are initialized in the up state. Units constrained to remain in the down state have their vertical displacement fixed throughout the simulation,
\[
v_\text{top}^i(t) = -2 h_0,
\]
while their horizontal motion remains unconstrained to preserve geometric compatibility and avoid over-constraining the structure. Constrained units are initialized in the down configuration, whereas unconstrained units are initialized in the up configuration, $v_\text{top}^i(0) = 0$.

We first use these simulations to calibrate $k_\theta$. A range of candidate $k_\theta$ values is tested under the same displacement-controlled loading protocol, and the resulting force–displacement curves are compared to experimental measurements obtained for a 9-unit chain in which all units are initially up except the central one, which transitions from down to up. The optimal value of $k_\theta$ is selected by minimizing the mean squared error (MSE) between simulation and experiment. Structure~\#a ($h = 0.7$~mm) and Structure~\#b ($h = 0.6$~mm) from Table~\ref{tab:geom} are analyzed to capture distinct torsional stiffness values. Using Eqs.~(\ref{eq:kl}) and~(\ref{eq:kg}), we obtain $k_\text{beam} = 21.1$~N/m and $k_\text{strut} = 3.5$~kN/m. A control stiffness of $k_{\mathrm{control}} = 1$~MN/m is adopted. The best-fit values are:
\[
k_\theta (h = 0.6~\mathrm{mm}) = 0.77~\mathrm{mN{\cdot}m/rad}, 
\quad 
k_\theta (h = 0.7~\mathrm{mm}) = 1.02~\mathrm{mN{\cdot}m/rad}.
\]

\begin{figure*}[!hpt]
\begin{center}
\includegraphics[width = 0.7\columnwidth]{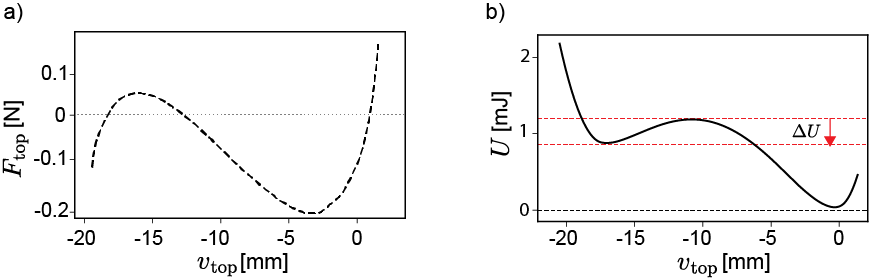} 
\caption{\label{fig:fit_f_energy} Quasi-static force–displacement response and reconstructed energy landscape for a wave-like configuration, for Structure \#b. \textbf{a)} Quasi-static force–displacement response of the central unit under asymmetric boundary conditions, with the left portion of the metamaterial constrained in the down state. \textbf{b)}  Effective energy landscape $U(v_\text{top})$ reconstructed by numerical integration of the force–displacement curve. The energy barrier $\Delta U$  is highlighted.}
\vspace{-15pt}
   \end{center}
\end{figure*}

Once calibrated, the same numerical framework and corresponding $k_\theta$ values are used to perform quasi-static force–displacement simulations for all configurations considered in Section~\ref{sec:SIforcedisp}. The experimental data and numerical predictions are reported in Fig.~\ref{fig:fit_f_energy}, including configurations in which all neighboring units are up, all are down, and mixed cases with one half up and the other half down.

Finally, we use this framework to systematically explore the influence of $k_{\text{beam}}$ and $k_\theta$ on the energy barrier $\Delta U$. As described in Section~\ref{sec:SIforcedisp}, we perform quasi-static simulations in which one side of the metamaterial is constrained in the down state while the other remains in the up state throughout loading. The quasi-static force–displacement response is obtained by recording the reaction force associated with the displacement-controlled actuation of the central unit (see Fig.~\ref{fig:fit_f_energy}a). An effective energy landscape is then reconstructed by numerical integration,
\begin{equation}
    U(v_\text{top}) = \int F(v_\text{top}) \,\mathrm{d}v,
\end{equation}
with the reference chosen such that $U(0) = 0$.

The resulting energy profile allows direct identification of the stationary points governing the transition. When present, the down configuration corresponds to a local minimum of $U$, the unstable intermediate configuration to a maximum, and the up configuration to a second minimum. We define $\Delta U$ as the energy barrier associated with the transition from the down to the up state when the von Mises trusses on one side of the central unit are maintained in the down state and those on the other side in the up state (see Fig.~\ref{fig:fit_f_energy}b). The predicted $\Delta U$ for structures \#a-d are reported in Table~\ref{tab:geom_energy_numerical}.

\begin{table}[h!]
    \centering
    \begin{tabular}{l|c|c}
        \toprule & \multicolumn{2}{c}{$\Delta U$ [mJ]} \\
        \textbf{Structure}  & \textbf{ Experimental } & \textbf{Numerical model} \\ 
        \midrule
        Structure \#a & 0.31 $\pm$ 0.21 & 0.19 \\
        Structure \#b & 0.34 $\pm$ 0.2 &  0.30\\
        Structure \#c & 0.37 $\pm$ 0.09 & 0.51 \\
        Structure \#d & 0.42 $\pm$ 0.05 & 0.62\\
        \bottomrule
    \end{tabular} 
    \caption{Summary $\Delta U$ experimental and numerical.}
    \label{tab:geom_energy_numerical}
\end{table}

\subsubsection{Assessing multistability}

In these simulations, we assume that at time $t=0$ all units are in their down configurations defined by first 
\begin{equation}
    v_\text{top}^{i}=-2 h_0, \quad u_{top}^i=u_{beam}^i=0\quad  \text{for } i=0,..., N
\end{equation}

To assess the stability of the down configuration, we then let the system relax without applying any external loading and monitor the displacements of the top masses. If all units remain in their down configuration during relaxation, we consider the configuration stable.

\subsubsection{Simulations of wave propagation}

In these simulations, we assume that at time $t=0$ all units are in their down configurations defined by first 
\begin{equation}
    v_\text{top}^{i}=-2 h_0, \quad u_{top}^i=u_{beam}^i=0\quad  \text{for } i=0,..., N
\end{equation}

To simulate wave propagation, we perturb the 
$i$-th unit at time $t=0$ by imposing 
\begin{equation}
    v_\text{top}^i(t=0)= -1~\mathrm{mm}
\end{equation} 
and then monitor the response of the structure over time. A summary of the predicted velocities for structures \#a-d is reported in Table~\ref{tab:geom_speed_numerical}.

\begin{table}[h!]
    \centering
    \begin{tabular}{l|c|c}
        \toprule
        & \multicolumn{2}{c}{$c_{\text{wave}}$ [units/s]} \\
        \textbf{Structure}  & \textbf{ Experimental } & \textbf{Numerical model} \\ 
        \midrule
        Structure \#a & 54.1 $\pm$ 3.2 & 53.7\\
        Structure \#b & 40.1 $\pm$ 3.9  & 38.8 \\
        Structure \#c & 37.2 $\pm$ 4.1 (partial, over 3-20 units) & 38.6 (partial, over 10 units)\\
        Structure \#d & No propagation  & No propagation \\
        \bottomrule
    \end{tabular}
    \caption{Summary of the speed $c_{wave}$.}
    \label{tab:geom_speed_numerical}
\end{table}

\clearpage 
\subsection{Influence of the base curvature on $c_{\text{wave}}$}\label{si:curvature}
\begin{figure}[!hpt]
\begin{center}
\includegraphics[width = 0.6\columnwidth]{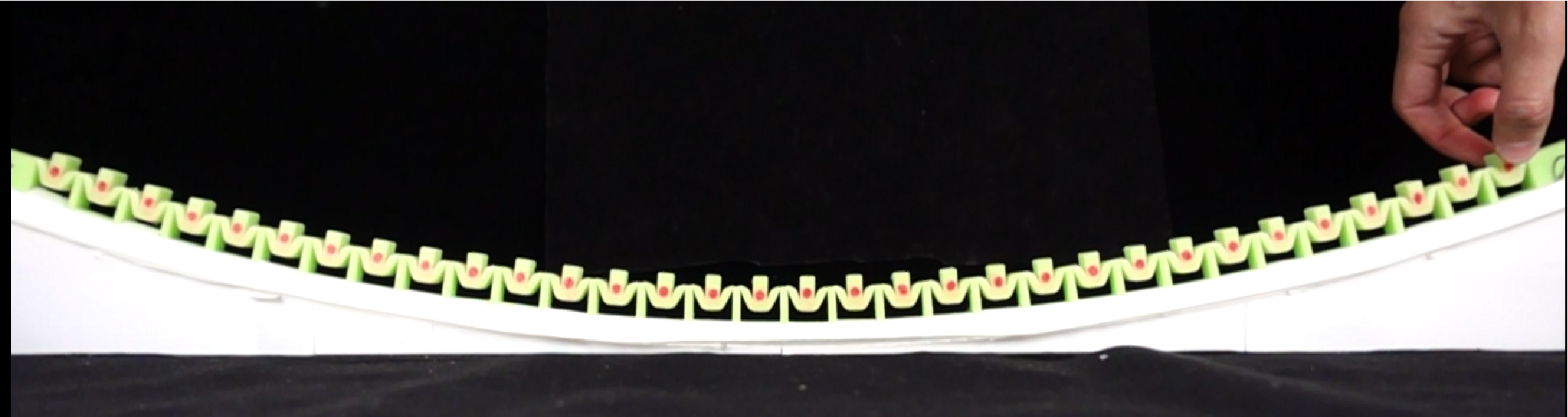} 
\caption{\label{fig:curvature_experim}  Dynamical test set-up at the variation of the base curvature. The structure is subject to constant base curvature $\kappa =-1$~m$^{-1}$.  }
\vspace{-15pt}
    \end{center}
\end{figure}

\begin{figure}[!hpt]
\begin{center}
\includegraphics[width = 0.7\columnwidth]{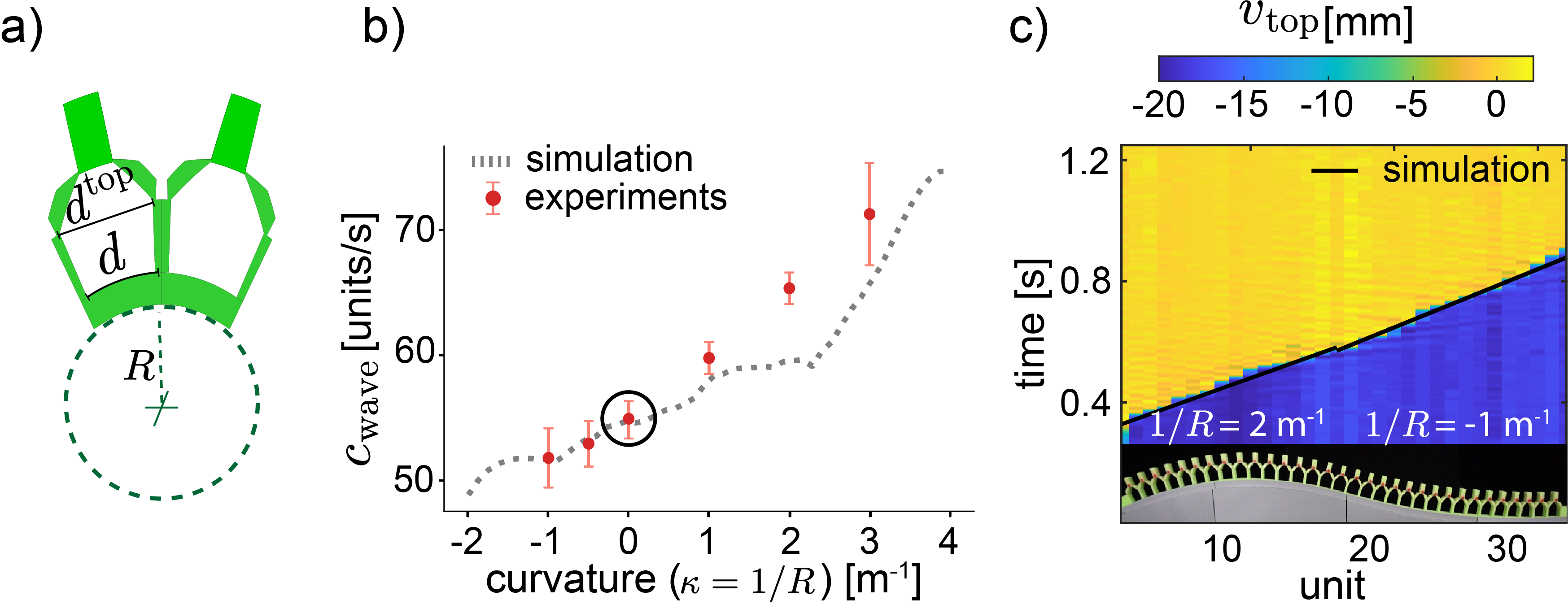} 
\caption{\label{fig:curvatureEffect} \textbf{Tuning the propagation velocity with $\kappa$.}  All structures have geometric parameters h = $0.7$~mm and w = $1.4$~mm. The reference experimental point is circled. \textbf{a)} Schematic illustrating the definition of the base curvature $\kappa = 1/R$ \textbf{b)} Velocity evolution with respect to added curvature $\kappa$ (simulated and experimental data) \textbf{c)} Spatiotemporal displacement diagram for a structure with two distinct curvature radii: $ \forall j \in [1, 16], \kappa = 2$~m$^{-1}$ (velocity $56.5$~units/s) and $ \forall j \in [17, 32], \kappa = -1$~m$^{-1}$ (velocity $51.5$~units/s) (see Supplementary Video 3).}
\vspace{-15pt}
    \end{center}
\end{figure}

In Fig.~4 of the main text, we demonstrate that the wave speed can be tuned on demand by varying the mass of the von Mises truss. Here, we show that a similar tuning effect can be achieved by modifying the base curvature. An example of experimental setup is illustrated in Fig.~\ref{fig:curvature_experim}. Imposing a curvature $\kappa$ at the base of the structure modifies the distance between the tops of the vertical beams, denoted $d^{\text{top}}$ (see Fig.~\ref{fig:curvatureEffect}a). For $\kappa > 0$, the beams bend farther apart (i.e., $d^{\text{top}} > d$), whereas for $\kappa < 0$ they bend toward each other ($d^{\text{top}} < d$). Importantly, the inclination angle of the von Mises truss, $\theta_0$, depends on this top-end spacing. In particular, positive curvature ($\kappa > 0$) reduces $\theta_0$ and lowers the energy barrier between the down and up states, thereby enabling faster wave propagation. Conversely, negative curvature increases this energy barrier and slows the propagating front.

We demonstrate this effect experimentally by clamping a 32-unit metamaterial onto bases with prescribed curvature $\kappa = 1/R \in [-1,\,-0.5,\,1,\,2,\,3]~\text{m}^{-1}$.  
Over this range, the measured wave speed varies by roughly $40~\%$, increasing from approximately $50$ to $70$~units/s as curvature changes from negative to positive values (Fig.~\ref{fig:curvatureEffect}b). We also investigate this effect with the numerical model, where the effect of curvature is captured by prescribing the corresponding change in the top-end spacing $d^\mathrm{top}$, which directly sets the prescribed inclination angle $\theta_0$ for each unit. Simulations reproduce the experimentally observed increase of $c_\mathrm{wave}$ with $\kappa$. Experimental results and model predictions show that when $|\kappa|$ is either too large or too small, the structure becomes fully monostable (the down state cannot be accessed) or fully bistable, respectively, thereby suppressing wave propagation. Furthermore, changing $\kappa$ on two halves of the 32-unit chain leads to a change in wave speed along the structure, from $c_{\text{wave}} = 56.5$~units/s in the left half of the structure with $\kappa =2$~m$^{-1}$ to $c_{\text{wave}} = 51.5$~units/s in the right half of the structure with $\kappa =-1$~m$^{-1}$ (Fig.~\ref{fig:curvatureEffect}c and Supplementary Video 3). This result demonstrates that local tuning of the base curvature distribution provides a means of spatially controlling wave propagation, albeit with a weaker effect than mass modulation (see Fig.~4 of the manuscript).

\clearpage

\subsection{Additional experimental results} 
~ 

\FloatBarrier

\begin{figure*}[!hpt]
\begin{center}
\includegraphics[width = 0.49\columnwidth]{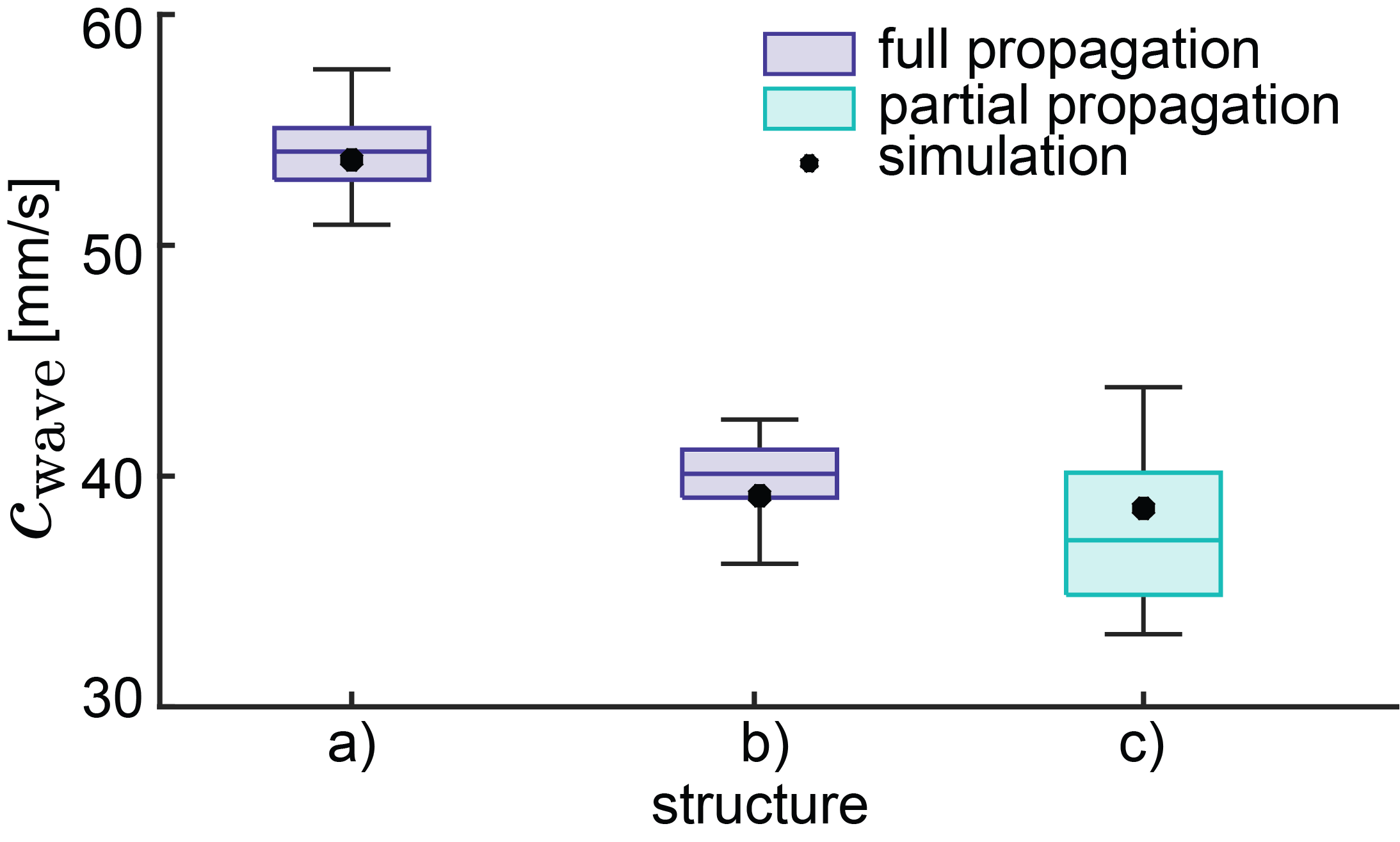} 
\caption{\label{fig:ModelValid} Comparison of experimentally measured and simulated propagation velocities. Simulated values are $53.7$~units/s for the reference ($h=0.7$~mm, $w_\text{beam}=1.4$~mm), $38.8$~units/s for the smaller-hinge case ($h=0.6$~mm), and $38.6$~units/s for the wider-beam case ($w_\text{beam}=1.7$~mm). }
\vspace{-15pt}
   \end{center}
\end{figure*}

\begin{figure*}[!hpt]
\begin{center}
\includegraphics[width = 0.99\columnwidth]{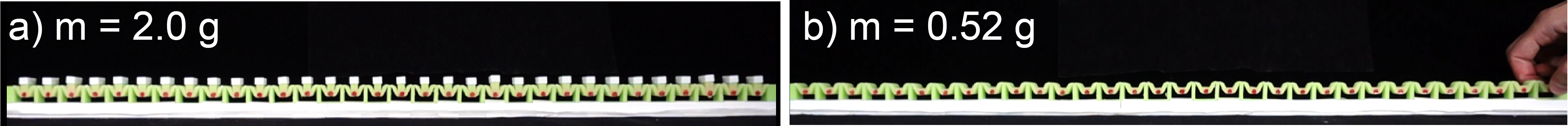} 
\caption{\label{fig:Mass}  Dynamical test set-up at the variation of top mass.  \textbf{a)} A structure with a top mass of $2.0$~g. Additional mass is added using small 3D-printed "hats" placed on top of each pull tab. \textbf{b)} Structure with a reduced top mass of $0.52$~g, achieved by removing material from the pull tab.}
\vspace{-15pt}
   \end{center}
\end{figure*}

\begin{figure*}[!hpt]
\begin{center}
\includegraphics[width = 0.99\columnwidth]{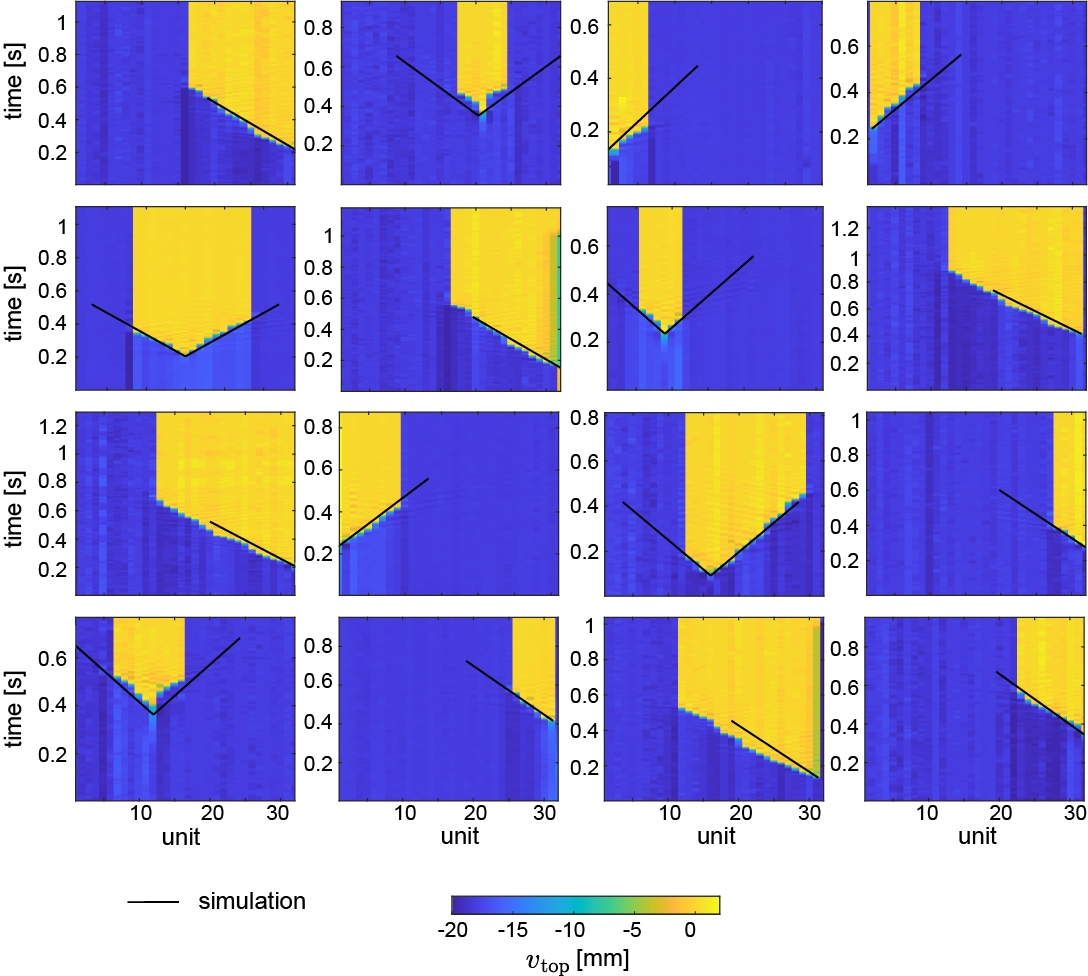} 
\caption{\label{fig:prop_partial}  Spatiotemporal displacement diagrams for structure~\#c. Each diagram represents a different experiment, showing the variability in the case of partial propagation. The simulated velocity curve is shown in black.}
\vspace{-15pt}
   \end{center}
\end{figure*}

\end{document}